\documentclass[12pt]{article}
\usepackage{hyperref,slashed,graphicx,color,amsmath,amssymb}
\usepackage[numbers,sort&compress]{natbib}
\usepackage{subfig}

\allowdisplaybreaks 

\begin{document}

\vskip 1cm
\begin{center}
{ \bf\large Perturbativity Limits for Scalar Minimal Dark Matter with Yukawa Interactions: Septuplet
 \vskip 0.2cm}
\vskip 0.3cm
{Chengfeng Cai$^a$, Ze-Min Huang$^a$, Zhaofeng Kang$^b$, Zhao-Huan Yu$^{a,c,d}$,\\
and Hong-Hao Zhang$^a$\footnote{Email: zhh98@mail.sysu.edu.cn}}
\vskip 0.3cm
{\it \small $^a$School of Physics and Engineering, Sun Yat-Sen University, Guangzhou 510275, China\\
$^b$School of Physics, Korea Institute for Advanced Study, Seoul 130-722, Korea\\
$^c$Key Laboratory of Particle Astrophysics,
Institute of High Energy Physics, Chinese Academy of Sciences,
Beijing 100049, China\\
$^d$ARC Centre of Excellence for Particle Physics at the Terascale,
School of Physics, The~University of Melbourne, Victoria 3010, Australia}
\end{center}
\vspace{0.3cm}
\begin{center}
{\bf Abstract}
\end{center}

The candidate of minimal dark matter (MDM) is limited if one demands perturbativity up to a very high scale, and it was believed that the MDM model with a real scalar septuplet could keep perturbative up to the Planck or GUT scale. In this work we point out that it is not true after taking into account the running of the quartic self-couplings of the scalar septuplet.
For the septuplet mass around $10$~TeV, which is suggested by the observed dark matter relic abundance, these couplings would hit the Landau pole at a scale $\sim 10^8-10^9$~GeV, much lower than the Planck scale.
We attempt to push up the Landau pole scale as high as possible by proposing an extension with extra Yukawa interactions of the septuplet. We find that in principle the Landau pole could be deferred to a scale of $\sim 10^{14}$~GeV if one could tolerate a serious fine-tuning of the initial condition of the Yukawa coupling.
Moreover, if the MDM particle mass could be relaxed to $\sim10^8$~GeV, which would need some nonthermal production mechanisms to give a correct relic abundance, the Landau pole scale could be pushed up above the Planck scale.

\newpage
\tableofcontents

\section{Introduction}

Among a large pool of dark matter (DM) models, the minimal dark matter (MDM) model~\cite{Cirelli:2005uq} is of special interest.
There are extensive studies on its implication for DM relic abundance~\cite{Cirelli:2005uq,Cirelli:2007xd,Cirelli:2009uv,Hambye:2009pw}
and its predictions for direct detection~\cite{Cirelli:2005uq,Cirelli:2007xd,Cirelli:2009uv,Hambye:2009pw,Ling:2009mu,Earl:2013jsa},
indirect detection~\cite{Cirelli:2005uq,Cirelli:2007xd,Cirelli:2008id,Cirelli:2008jk,Cirelli:2009uv,Hambye:2009pw,Ling:2009mu,Cirelli:2015bda,Garcia-Cely:2015dda},
and collider~\cite{Cirelli:2005uq,Cirelli:2009uv,Buckley:2009kv,Cai:2012kt,Earl:2013jsa,Cirelli:2014dsa,Harigaya:2015yaa,Ostdiek:2015aga} experiments.
A MDM particle is the electrically neutral component of a $SU(2)_L\times U(1)_Y$ multiplet in a high dimensional representation, which is denoted as $(2j+1,Q_Y)$ with $j$ the half integer and $Q_Y$ the hyper charge. Minimally, MDM annihilations only involve gauge interactions. As a consequence, the MDM particle mass can be predicted from the observed DM abundance. Moreover, it may shed light on the mechanism of DM stability by virtue of an accidental symmetry rather than an artificial protecting symmetry. As long as the dimension of the representation is sufficiently high, the electroweak gauge symmetries could forbid any coupling between the MDM and the standard model (SM) particles which can lead to the MDM decay, at tree or even at nonrenormalizable level. For instance, a fermonic MDM particle in $(5,0)$ can be stable up to dimension-6 operators. The quintuplet MDM has been well studied in Refs.~\cite{Cirelli:2005uq,Cirelli:2007xd,Cirelli:2009uv,Cirelli:2015bda,Ostdiek:2015aga,Garcia-Cely:2015dda}.

Another option is a real scalar septuplet $(7,0)$~\footnote{Recently, a dimension-5 operator $\sim\Phi^3 H^\dagger H$ that violates the accidental $Z_2$ symmetry was pointed out by Ref.~\cite{DiLuzio:2015oha}. But this operator only induces DM decay at loop level.} which, to our knowledge, was less studied~\cite{Hambye:2009pw,Ling:2009mu,Cai:2012kt,Earl:2013jsa,Garcia-Cely:2015dda}, in particular on its self-consistency in the framework of quantum field theory.
The observed DM relic abundance suggests that this septuplet MDM particle has a threshold mass of $\sim 8$~TeV (or $\sim 22$~TeV) without (or with) including the Sommerfeld enhancement (SE) effect~\cite{Hambye:2009pw,Ling:2009mu}. In this paper we revisit this DM candidate in light of some recent progresses, in particular the work in Ref.~\cite{Hamada:2015bra}, which for the first time systematically calculated the beta functions of quartic couplings of MDM and found out that they have deep implications to the perturbativity of the model. Previously, the perturbativity was only checked with respect to gauge couplings, concretely $g_2$, and it was claimed the septuplet MDM model can keep perturbative up to the Planck scale~\cite{Cirelli:2005uq} (but it was recently lowered down to the GUT scale after including two-loop contributions~\cite{DiLuzio:2015oha}). Nevertheless, a renormalization group equation (RGE) study on the MDM quartic self-interaction couplings shows that they will hit the Landau pole (LP) at a scale merely around $10^8$ GeV~\cite{Hamada:2015bra}, much faster than gauge couplings.

In this article we attempt to defer the appearance of the Landau pole by introducing sizable Yukawa interactions of the septuplet MDM.
It is found that the most helpful case is the existence of a coupling of the scalar septuplet to a fermion triplet and a fermion quintuplet. If the two fermion multiplets are active around 100 TeV, the perturbativity can be kept up to around $\Lambda_{\rm LP}\simeq$ 10$^{14}$ GeV. As a bonus of our scheme, the fermion triplet can be used to explain neutrino mass origin via the type-III seesaw mechanism\footnote{Of course, the realistic neutrino mixings require at least two triplets, but the other one is assumed to be very heavy for the sake of a Landau pole of $g_2$ as high as possible.}~\cite{Foot:1988aq}.
Careful studies, both numerical and analytical, show that it is at the price of serious fine-tuning on the initial conditions. Relaxing the MDM particle mass to $10^8$ GeV, under the assumption that the observed relic density is obtained in other ways, the model can even be perturbative near the Planck scale.

Other aspects of the septuplet MDM model are also investigated, and some of them are new. Different to fermionic MDM, scalar MDM couplings to the Higgs field are always allowed. If these couplings are significant they may alter physics associating with the Higgs, e.g., electroweak vacuum stability. We have to guarantee that the electroweak vacuum is  indeed the global minimum in the presence of a scalar septuplet. Thus we figure out the vacuum stability (VS) conditions of the whole scalar potential, which are non-trivial owing to the complicated quartic couplings of the septuplet. It is found that the electroweak vacuum can be absolutely stable at any scale below $\Lambda_{\rm LP}$.

We organize the paper as follows. In Sec.~\ref{sect2} we introduce the septuplet MDM model and study several relevant phenomenologies. In Sec.~\ref{sect3} we investigate the perturbativity bound on the septuplet model as well as its extension with Yukawa interaction. Sec.~\ref{sect4} gives our conclusions and discussions.

\section{The Real Scalar Septuplet MDM Model}\label{sect2}

In this section we begin by describing the model. We pay special attention to the scalar potential which is overlooked before. Then we investigate the perturbativity bound on the model including the evolution of quartic couplings of MDM self-interactions.

\subsection{Details of the septuplet model}

The MDM model was firstly proposed in Ref.~\cite{Cirelli:2005uq}.
The idea is to extend the SM with a colorless $SU(2)_L$ multiplet, whose electrically neutral component plays the role of the DM candidate.
If the multiplet belongs to a representation with sufficiently high dimension, it would be unable to construct any renormalizable decay operator for this multiplet.
Consequently, its neutral component could be stable and weakly couple to other particles.

Taking into account the nonrenormalizable operators, the DM stability condition sets a lower bound on the dimension of the representation $n$. The bound is $n\geq 5$ for fermionic multiplets and $n\geq 7$ for scalar multiplets.
On the other hand, an upper bound on $n$ can be determined by the perturbativity of the $SU(2)_L$ gauge coupling $g_2$.
It requires $n\leq 5$ for Majorana fermionic multiplets and $n\leq 8$ for scalar multiplets.
Therefore, the minimal choice is a Majorana fermionic quintuplet with a hypercharge $Y=0$, and the next-to-minimal extension is a real scalar septuplet with $Y=0$.
Note that for Dirac fermionic and complex scalar multiplets, $Y\neq 0$ leads to the spin-independent DM-nuclei scattering through the $Z$ boson exchange and can be easily excluded by current direct detection experiments.

General discussions for a model with an extra $SU(2)_L$ scalar multiplet whose neutral component is a DM candidate can be found in Ref.~\cite{Hambye:2009pw}.
Here we briefly revisit the real scalar septuplet case.
The scalar septuplet $\Phi$ can be expressed as
\begin{eqnarray}
\Phi=\frac{1}{\sqrt{2}}(\Delta^{(3)},\ \Delta^{(2)},\ \Delta^{(1)},\ \Delta^{(0)},\ \Delta^{(-1)},\ \Delta^{(-2)},\ \Delta^{(-3)})^\mathrm{T},
\end{eqnarray}
where $(\Delta^{(Q)})^\ast=\Delta^{(-Q)}$.
The gauge covariant derivative of the septuplet with $Y=0$ is
\begin{eqnarray}
D_\mu \Phi=\partial_\mu \Phi-ig_2W^a_\mu\tau^a \Phi,
\end{eqnarray}
where $\tau^a$ is the $SU(2)$ generators in the 7-dimensional representation. They satisfy the $su(2)$ algebra, i.e., $[\tau^a,\tau^b]=i\epsilon^{abc}\tau^c$.
As usual, we choose the spherical basis and $\tau^3=\textrm{diag}\{3,2,1,0,-1,-2,-3\}$.
It is convenient to define the ladder operators $\tau^\pm=\tau^1\pm i\tau^2$.
Since $W_\mu^\pm\equiv (W_\mu^1\mp iW_\mu^2)/\sqrt{2}$ and $W_\mu^3 = \sin\theta_W A_\mu + \cos\theta_W Z_\mu$ with $\theta_W$ denoting the Weinberg angle, we have
\begin{equation}
W_\mu^a\tau^a=(s_W A_\mu + c_W Z_\mu)\tau^3+\frac{1}{\sqrt{2}}(W_\mu^+\tau^+ +W_\mu^-\tau^-),
\end{equation}
where $s_W\equiv\sin\theta_W$ and $c_W\equiv\cos\theta_W$.
The commutators of these operators can be easily derived:
\begin{eqnarray}
[\tau^+,\tau^-]=2\tau^3,\ [\tau^3,\tau^\pm]=\pm\tau^\pm.
\end{eqnarray}

Since all $SU(2)$ representations are pseudo-real, we can relate an $n$-dimensional representation to its complex conjugate by a transformation matrix $T_{(n)}$:
\begin{eqnarray}
T_{(n)}\tau^a T_{(n)}^{-1}=-(\tau^a)^\ast.
\end{eqnarray}
For the spherical basis $|e_k^{(n)}\rangle$ (eigenstates of $\tau^3$ in the $n$-dimensional representation), $T_{(n)}$ satisfies
\begin{eqnarray}
T_{(n)}|e_k^{(n)}\rangle=\left\{\begin{matrix}(-)^{n+1}|e_k^{(n)}\rangle,&k\geq0;\\
|e_{-k}^{(n)}\rangle,&k<0.\end{matrix}\right.
\end{eqnarray}
Thus $T_{(7)}$ is a $7\times 7$ matrix whose minor diagonal elements are 1 and the rest elements are 0.
The conjugate of $\Phi$ can be constructed as $\tilde{\Phi}=T_{(7)}\Phi^*$.
Particularly, for a real scalar septuplet, $\tilde{\Phi} = \Phi$,
which is consistent with the condition $(\Delta^{(Q)})^\ast=\Delta^{(-Q)}$.

For an $n$-dimensional representation, the ladder operator $\tau^+$ satisfies
\begin{eqnarray}
\tau^+|e_k^{(n)}\rangle=\left\{\begin{matrix}-\sqrt{(j-k)(j+k+1)}|e_{k+1}^{(n)}\rangle,&k\geq0;\\
\sqrt{(j-k)(j+k+1)}|e_{k+1}^{(n)}\rangle, &k<0.\end{matrix}\right.
\end{eqnarray}
Here $j=(n-1)/2$, and $k=-j,-j+1,\cdots,j-1,j$.
For $n=2j+1=7$, $\tau^\pm$ can be explicitly expressed as
\begin{equation}
\begin{split}
\tau^+=\begin{pmatrix}0&-\sqrt{6}&&&&&\\&0&-\sqrt{10}&&&&\\&&0&-\sqrt{12}&&&\\&&&0&\sqrt{12}&&\\&&&&0&\sqrt{10}&\\&&&&&0&\sqrt{6}\\&&&&&&0\end{pmatrix},
\quad \tau^-=(\tau^+)^\mathrm{T}.
\end{split}
\end{equation}
With these matrices, the kinetic term and the couplings to the gauge fields of each component can be explicitly written down as
\begin{eqnarray}\label{gaugecouple}
\mathcal{L}_1&=&(D_\mu \Phi)^\dag D^\mu \Phi
\nonumber\\
&=&\frac{1}{2}(\partial_\mu\Delta^{(0)})^2+\sum_{Q=1}^3(\partial_\mu\Delta^{(Q)})(\partial^\mu\Delta^{(-Q)})
+\sum_{Q=1}^3(QeA^\mu+Qg_2c_W Z^\mu)\Delta^{(-Q)}i\overleftrightarrow{\partial_\mu}\Delta^{(Q)}\nonumber\\
&&-g_2W^{+,\mu}(\sqrt{3}\Delta^{(-3)}i\overleftrightarrow{\partial_\mu}\Delta^{(2)}
+\sqrt{5}\Delta^{(-2)}i\overleftrightarrow{\partial_\mu}\Delta^{(1)}
+\sqrt{6}\Delta^{(-1)}i\overleftrightarrow{\partial_\mu}\Delta^{(0)})
\nonumber\\
&&-g_2W^{-,\mu}(\sqrt{3}\Delta^{(-2)}i\overleftrightarrow{\partial_\mu}\Delta^{(3)}
+\sqrt{5}\Delta^{(-1)}i\overleftrightarrow{\partial_\mu}\Delta^{(2)}
+\sqrt{6}\Delta^{(0)}i\overleftrightarrow{\partial_\mu}\Delta^{(1)})\nonumber\\
&&+(e^2A_\mu A^\mu+g_2^2c_W^2 Z_\mu Z^\mu+2eg_2c_W A_\mu Z^\mu)
\sum_{Q=1}^3Q^2\Delta^{(Q)}\Delta^{(-Q)}
\nonumber\\
&&+g_2^2 W^+_\mu W^{-,\mu}[6(\Delta^{(0)})^2+11\Delta^{(1)}\Delta^{(-1)}
+8\Delta^{(2)}\Delta^{(-2)}+3\Delta^{(3)}\Delta^{(-3)}]
\nonumber\\
&&-g_2^2\big\{ W^+_\mu(s_W A^\mu + c_W Z^\mu)
(\sqrt{6}\Delta^{(0)}\Delta^{(-1)}+3\sqrt{5}\Delta^{(1)}\Delta^{(-2)}+5\sqrt{3}\Delta^{(2)}\Delta^{(-3)})
\nonumber\\
&&+ W^+_\mu W^{+,\mu} [3(\Delta^{(-1)})^2
-\sqrt{30}\Delta^{(0)}\Delta^{(-2)}
-\sqrt{15}\Delta^{(1)}\Delta^{(-3)}]
+\mathrm{h.c.}\big\},
\end{eqnarray}
where $\overleftrightarrow{\partial_\mu}$ is defined as $F\overleftrightarrow{\partial_\mu} G=F\partial_\mu G-G\partial_\mu F$.

In this model, the potential is not only constructed by the Higgs doublet $H$, but also constructed by the septuplet $\Phi$.
Since $\Phi$ is real, the operator $\Phi^\dag\tau^a\Phi$ vanishes, but a term as $(\Phi^\dag T^aT^b\Phi)^2$ is allowed, and the general form of the potential is
\begin{eqnarray}\label{potential}
V&=&\mu^2 H^\dag H+m^2\Phi^\dag\Phi+\lambda(H^\dag H)^2+\lambda_2(\Phi^\dag\Phi)^2\nonumber\\
&&+\lambda_3(H^\dag H)(\Phi^\dag\Phi)+\frac{\lambda_4}{48}(\Phi^\dag T^aT^b\Phi)^2.
\end{eqnarray}
Comparing with the SM, there are three more couplings $\lambda_2$, $\lambda_3$, and $\lambda_4$, and one more mass parameter $m$. 
Note that the last term in \eqref{potential} can be separated from the traceless part (with respect to the indices $a,b$) as follows
\begin{eqnarray}
\Phi^\dag T^aT^b\Phi=\frac{1}{2}\Phi^\dag\{T^a,T^b\}\Phi=\Phi^\dag
\left(\frac{1}{2}\{T^a,T^b\}-4\delta^{ab}\right)\Phi+4\delta^{ab}\Phi^\dag\Phi.
\end{eqnarray}
where we have used $\Phi^\dag T^a\Phi=0$ and $T^aT^a=C_2(j)\mathbf{1}=j(j+1)\mathbf{1}$ with $j=3$.
Define $(S^{ab})_{ij}=\frac{1}{2}\{T^a,T^b\}_{ij}-4\delta^{ab}\delta_{ij}$, and $(\Phi^\dag T^aT^b\Phi)^2$ can be rewritten as a sum of two quadratic terms:
\begin{eqnarray}
(\Phi^\dag T^aT^b\Phi)^2=(\Phi^\dag S^{ab}\Phi)^2+48(\Phi^\dag\Phi)^2.
\end{eqnarray}
Thus the potential can be expressed as
\begin{eqnarray}\label{potential2}
V&=&\mu^2 H^\dag H+m^2\Phi^\dag\Phi+\lambda(H^\dag H)^2+(\lambda_2+\lambda_4)(\Phi^\dag\Phi)^2\nonumber\\
&&+\lambda_3(H^\dag H)(\Phi^\dag\Phi)+\frac{\lambda_4}{48}(\Phi^\dag S^{ab}\Phi)^2.
\end{eqnarray}

We assume that the vacuum expectation value (VEV) of the Higgs field is nonzero but the VEV of $\Phi$ remains zero.
Then the minimization condition implies that $\mu^2<0$ and $m^2-{\lambda_3\mu^2}/{(2\lambda)}\geq0$.
As in the SM, the VEV of the Higgs doublet is $\langle H\rangle=(0,{v}/{\sqrt{2}})^\mathrm{T}$, where $v\equiv\sqrt{{-\mu^2}/{\lambda}}=246.22$~GeV.

After the Higgs field acquires a VEV, the potential term $\lambda_3(H^\dag H)(\Phi^\dag\Phi)$ contributes to the masses of all components of $\Phi$. Even so, they are totally degenerate at the tree level with a value of $m_0$, which satisfies
\begin{eqnarray}
m_{0}^2=m^2+\frac{\lambda_3v^2}{2}.
\end{eqnarray}
Mass splittings among the components are induced by loop corrections involving gauge bosons, and the charged components are slightly heavier than the neutral component.
For $m_0 \gg m_Z$, the mass splitting between $\Delta^{(Q)}$ and $\Delta^{(0)}$ is~\cite{Cirelli:2005uq}
\begin{equation}\label{eq:m_corr}
m_Q - m_0 = Q^2 \Delta m,
\end{equation}
where $\Delta m=\alpha_2m_W\sin^2({\theta_W}/{2})\simeq 167~\mathrm{MeV}$.
For $m_0\sim\mathcal{O}(\mathrm{TeV})$, these splittings are very tiny and we can still regard the components degenerate.

\subsection{Several relevant phenomenologies}

\subsubsection{DM phenomenologies}\label{relic}

The septuplet mass threshold affects the cosmological DM relic abundance. In the following, we will quickly review the relic abundance calculation and obtain the mass threshold favored by observation.
As discussed in Ref.~\cite{Hambye:2009pw},
the thermal DM relic abundance for a real scalar MDM model can be approximately expressed as
\begin{eqnarray}
\Omega_\mathrm{DM}h^2\simeq\frac{1.07\times10^9~\mathrm{GeV}^{-1}}{J(x_F)\sqrt{g_\ast}M_\mathrm{Pl}}~~
\text{with}~~J(x_F)=\int_{x_F}^\infty\frac{\langle\sigma_\mathrm{eff}v\rangle}{x^2}dx,
\end{eqnarray}
where $M_\mathrm{Pl}$ is the Planck mass and $g_*$ is the total number of effectively relativistic degrees of freedom.
The effective thermally averaged annihilation cross section is defined by
\begin{equation}
\langle\sigma_\mathrm{eff}v\rangle \equiv
\sum_{QQ'}\langle\sigma_{QQ'}v\rangle
\frac{n_Q^\mathrm{eq} n_{Q'}^\mathrm{eq}}{(n^\mathrm{eq})^2},
\end{equation}
where $\sigma_{QQ'}$ is the annihilation cross section between the multiplet components $\Delta^{(Q)}$ and $\Delta^{(Q')}$.
$n_Q^\mathrm{eq}=(m_QT/2\pi)^{3/2}\exp(-m_Q/T)$ is the thermal equilibrium density of $\Delta^{(Q)}$, and $n^\mathrm{eq}\equiv\sum_Q n_Q^\mathrm{eq}$.
By such a definition of $\langle\sigma_\mathrm{eff}v\rangle$, we take into account the coannihilation effect among the multiplet components.
$x_F$ is the freeze-out parameter that can be obtained by solving the equation
\begin{eqnarray}
x_F=\ln\frac{0.0038M_\mathrm{Pl}g_\mathrm{eff}m_0\langle\sigma_\mathrm{eff}v\rangle}{\sqrt{g_\ast x_F}},
\end{eqnarray}
where the effective number of degrees of freedom $g_\mathrm{eff}=\sum_Q n_Q^\mathrm{eq}/n_0^\mathrm{eq}$.

Neglecting the mass splittings among the multiplet components, we have $n_Q^\mathrm{eq}\simeq n_{Q'}^\mathrm{eq}$ for any $\Delta^{(Q)}$ and $\Delta^{(Q')}$, and hence $\langle\sigma_\mathrm{eff}v\rangle \simeq \sum_{QQ'}\langle\sigma_{QQ'}v\rangle/n^2$ and $g_\mathrm{eff}\simeq n$.
Since the $s$-wave annihilations into gauge and Higgs bosons are dominant, for $m_0\gg m_h$ we have
\begin{equation}
\langle\sigma_\mathrm{eff}v\rangle \simeq \frac{(n^2-1)(n^2-3)}{n}\frac{\pi\alpha_2^2}{8m_0^2}
+ \frac{1}{n}\frac{\lambda_3^2}{16\pi m_0^2},
\end{equation}
where $\alpha_2\equiv g_2^2/(4\pi)$.
For a heavier DM particle, $\langle\sigma_\mathrm{eff}v\rangle$ would be smaller and lead to a larger abundance.
For the septuplet model, $n=7$, and we take $x_F\simeq25$ and $\sqrt{g_\ast}\simeq10.33$ for $T\sim\mathcal{O}(\mathrm{TeV})$,
and calculate the relic abundance.

\begin{figure}[htbp]
\centering
\includegraphics[width=0.6\textwidth]{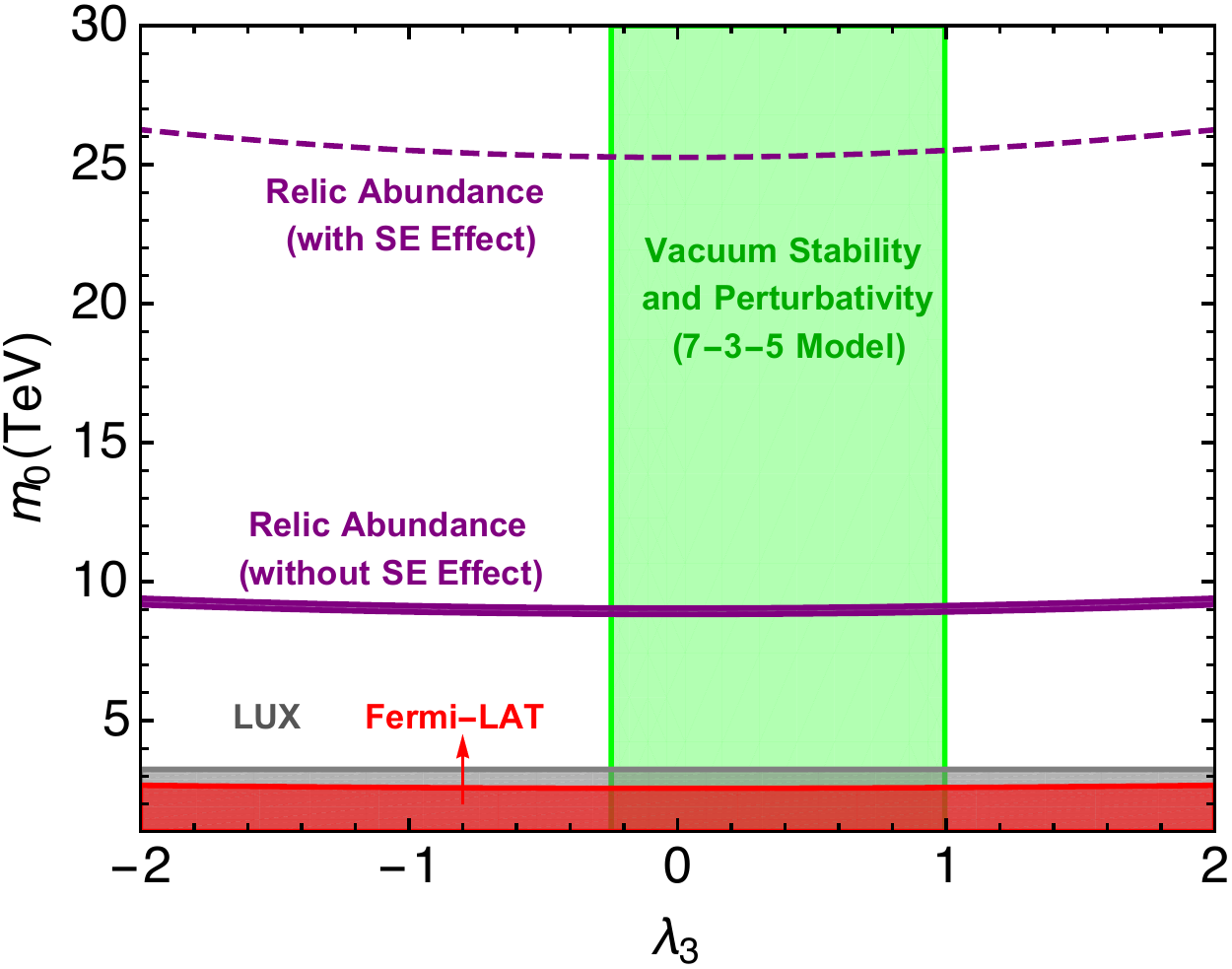}
\caption{Regions favored by the observed DM relic abundance in the $\lambda_3$-$m_0$ plane.
The purple (dashed) strip corresponds to the $1\sigma$ range of the relic abundance measured by the Planck experiment for the case without (with) the Sommerfeld enhancement effect. The green band is the region satisfying the vacuum stability and the perturbativity conditions in the 7-3-5 model.
The grey and red regions are excluded by the LUX and Fermi-LAT results at 90\% and 95\% C.L., respectively.}\label{m0l3}
\end{figure}

Fixing the relic abundance to its observed value $\Omega_\mathrm{DM}h^2=0.1196\pm0.0031$~\cite{Planck:2015xua},
we can constrain the parameters $m_0$ and $\lambda_3$.
In Fig.~\ref{m0l3}, the $1\sigma$ favored region is denoted by a purple strip in the $\lambda_3$-$m_0$ plane.
In this strip, $m_0$ increases as $|\lambda_3|$ increases.
For $\lambda_3=0$, $m_0$ achieves its threshold mass, $\simeq 8.8~\mathrm{TeV}$.
The calculation above has not included the SE effect, which can increase annihilation cross sections and hence reduce the relic abundance for fixed $m_0$ and $\lambda_3$.
As suggested in Refs.~\cite{Cirelli:2007xd,Hambye:2009pw},
the SE factor at the freeze-out epoch is almost a constant, and we can simply increase $\langle\sigma_\mathrm{eff}v\rangle$ by a scale factor $\simeq 8$ to take this effect into account in the septuplet model.
The dashed purple line in Fig.~\ref{m0l3} corresponds to the observed relic abundance including the SE effect.
In this case, the threshold mass is about 25~TeV.

A more accurate treatment for the DM relic abundance in real scalar MDM models can be found in Ref.~\cite{Hambye:2009pw},
and gives a threshold septuplet mass of  $7.9~(22.4)~\textrm{TeV}$ without (with) the SE effect using the WMAP data. These results are about 10\% smaller than our approximate results here.
Nonetheless, these discrepancies would not essentially change the following analysis on the vacuum stability and the perturbativity of couplings.

We also briefly comment on the direct and indirect detection bounds on this DM candidate. At tree level, only the SM Higgs boson mediates the DM-nucleon scattering, but the rate is greatly suppressed by the heavy DM mass squared and thus is negligible as long as $\lambda_3$ is not very large (later we will see that this is true after taking into account vacuum stability). However, at loop level there is a contribution that is not so suppressed~\cite{Cirelli:2005uq}:
\begin{equation}
\sigma^{(1)}_{\text{DM-}N}=\frac{36\pi\alpha_2^4 f_N^2 m_N^4}{m_W^2}\left(\frac{1}{m_W^2}+\frac{1}{m_h^2}\right)^2.
\end{equation}
where $f_N \simeq 0.3$ is the nucleonic form factor~\cite{Ellis:2000ds} and $m_N\simeq 1~\mathrm{GeV}$ is the nucleon mass. This cross section, independent of $m_0$, has a value of $\simeq 4\times10^{-44}~\textrm{cm}^2$. Therefore, the 90\% C.L. exclusion limit from LUX~\cite{Akerib:2013tjd} can exclude the range of $m_0\lesssim 3.3~\mathrm{TeV}$, as shown by the grey region in Fig.~\ref{m0l3}. For indirect searches, the most promising annihilation channel is the $W^+W^-$ channel, which has a cross section $\sim 9g_2^4/\pi m_{\rm DM}^2$~\footnote{The $ZZ/hh$ channels produce similar gamma-ray spectra to that from the $WW$ channel~\cite{Cirelli:2010xx}, but the cross section $\simeq \lambda_3^2/64\pi m_{\rm DM}^2$, depending on $\lambda_3$.}. The Fermi-LAT limit~\cite{Ackermann:2015zua} excludes the range of $m_0\gtrsim 2.5~\mathrm{TeV}$ for $|\lambda_3| \lesssim 2$, as shown by the red region in Fig.~\ref{m0l3}.

\subsubsection{Vacum stability}

The VS conditions can be obtained by means of the copositive criteria~\cite{Kannike:2012pe}. However, it is not quite straightforward to obtain these conditions in the septuplet model. The obstacle is from the quartic term $\mathcal{Q}_1\equiv (\Phi^\dag S^{ab}\Phi)^2$, which, unlike the conventional term $\mathcal{Q}_0\equiv (\Phi^\dag\Phi)^2$, yields non-universal (even in sign) quartic terms for different components. To see this, we explicitly expand $\mathcal{Q}_1$ in components as
\begin{eqnarray}
\frac{\mathcal{Q}_1}{48}=&&\frac{1}{8}(\Delta^{(0)})^4+\frac{21}{32}|\Delta^{(1)}|^4+\frac{25}{32}|\Delta^{(3)}|^4\nonumber\\
&&+\frac{1}{2} |\Delta^{(1)}|^2(\Delta^{(0)})^2+\frac{5}{4}|\Delta^{(2)}|^2(\Delta^{(0)})^2-\frac{5}{8}|\Delta^{(3)}|^2(\Delta^{(0)})^2\nonumber\\
&&+\frac{15}{16} |\Delta^{(1)}|^2 |\Delta^{(2)}|^2-\frac{5}{16}|\Delta^{(1)}|^2|\Delta^{(3)}|^2+\frac{25}{16} |\Delta^{(2)}|^2|\Delta^{(3)}|^2\nonumber\\
&&-\frac{\sqrt{15}}{8\sqrt{2}}(\Delta^{(-2)}(\Delta^{(1)})^2 \Delta^{(0)}+\Delta^{(2)}(\Delta^{(-1)})^2\Delta^{(0)})\nonumber\\
&&-\frac{\sqrt{15}}{8} (\Delta^{(-3)}(\Delta^{(1)})^3+(\Delta^{(-1)})^3\Delta^{(3)})\nonumber\\
&&+\frac{5}{16} \sqrt{15} (\Delta^{(-3)} \Delta^{(-1)}(\Delta^{(2)})^2+(\Delta^{(-2)})^2\Delta^{(1)}\Delta^{(3)})\nonumber\\
&&+\frac{15}{8 \sqrt{2}}(\Delta^{(-3)}\Delta^{(1)}\Delta^{(2)}\Delta^{(0)}+\Delta^{(-2)}\Delta^{(-1)}\Delta^{(3)}\Delta^{(0)}).
\end{eqnarray}
Vacuum stability concerns the behavior of the potential as the field values go to infinity. This allows us to reduce $\mathcal{Q}_1$. The key observation is that the ratio $\mathcal{Q}_1/(48\mathcal{Q}_0)$ reaches its maximum $25/32$ and minimum 0 when $\Phi$ goes to infinity along the $\Delta^{(3)}$ and $\Delta^{(2)}$ directions, respectively. Therefore, we can parametrize $\mathcal{Q}_1/48$ as $25 \rho \mathcal{Q}_0/32$ with $0\leq \rho \leq1$. Then the potential becomes
\begin{eqnarray}\label{potential3}
V&=&\mu^2 H^\dag H+m^2\Phi^\dag\Phi+\lambda(H^\dag H)^2+\lambda_3(H^\dag H)(\Phi^\dag\Phi)\nonumber\\
&&+\left[(\lambda_2+\lambda_4)+\frac{25}{32}\rho \lambda_4\right](\Phi^\dag\Phi)^2.
\end{eqnarray}
Now we can use the copositive criteria to get the VS conditions, which depend on the sign of $\lambda_4$. For $\lambda_4\geq0$, the bottom of the potential is achieved when $\rho=0$ and the VS conditions are
\begin{eqnarray}\label{VS1}
\left\{\begin{matrix}
\lambda\geq0,\\
\lambda_2+\lambda_4\geq0,\\
\lambda_3+2\sqrt{\lambda(\lambda_2+\lambda_4)}\geq0.
\end{matrix}\right.
\end{eqnarray}
While for $\lambda_4<0$ the bottom is achieved when $\rho=1$, and the VS conditions turn out to be
\begin{eqnarray}\label{VS2}
\left\{\begin{matrix}
\lambda\geq0,\\
\lambda_2+\frac{57}{32}\lambda_4\geq0,\\
\lambda_3+2\sqrt{\lambda\left(\lambda_2+\frac{57}{32}\lambda_4\right)}\geq0.
\end{matrix}\right.
\end{eqnarray}
Note that $\lambda$ can never be negative.

\section{Confronting perturbativity}
\label{sect3}

In the framework of quantum field theory, perturbativity is important to guarantee the self-consistency of perturbative calculations. The breakdown of perturbativity at some scale, known as the Landau pole scale $\Lambda_\mathrm{LP}$, implies that a new theory should appear hereafter. Perturbativity imposes a strong bound on models. In this section we firstly show that the minimal model suffers the Landau pole problem around $10^{8}$~GeV.
Then we attempt to push it up to $10^{14}$ GeV by introducing Yukawa couplings to the septuplet.

\subsection{Perturbativity bound on the septuplet model}

In the real scalar septuplet MDM model, there are mainly two modifications that may endanger perturbativity. One is a large positive contribution to the beta function of $g_2$ from the 7-dimensional representation. It drives $g_2$ towards the Landau pole more quickly. The other one is the focus of this paper, the fast increase of MDM self-couplings owing to their impressively large beta functions. Explicitly, the one-loop formulas can be calculated using the general formulas presented in Ref.~\cite{Cheng:1973nv} (or instead formulas up to two-loop are available using the code  \texttt{PyR@TE}~\cite{Lyonnet:2013dna}),
\begin{eqnarray}
\beta_{g_1}&=&\beta_{g_1}^\mathrm{SM},~~
\beta_{g_2}=\beta_{g_2}^\mathrm{SM}+\frac{1}{16\pi^2}\frac{14}{3}g_2^3,~~
\beta_{g_3}=\beta_{g_3}^\mathrm{SM},~~
\beta_{y_t}=\beta_{y_t}^\mathrm{SM},
\label{betafunc_1}\\
\beta_{\lambda}&=&\beta_{\lambda}^\mathrm{SM}+\frac{1}{16\pi^2}\frac{7}{2}\lambda_3^2,~~
\beta_{\lambda_2}=\frac{1}{16\pi^2}[30\lambda_2^2+2\lambda_3^2+\frac{45}{2}\lambda_4^2+51\lambda_2\lambda_4-144g_2^2\lambda_2],
\label{betafunc_2}\\
\beta_{\lambda_3}&=&\frac{1}{16\pi^2}\left[12\lambda\lambda_3+18\lambda_2\lambda_3+4\lambda_3^2+\frac{51}{2}\lambda_3\lambda_4+36g_2^4-\lambda_3\left(\frac{153}{2}g_2^2+\frac{9}{10}g_1^2-6y_t^2\right)\right],
\label{betafunc_3}\nonumber\\*
\\
\beta_{\lambda_4}&=&\frac{1}{16\pi^2}\left[288g_2^4+\frac{255}{8}\lambda_4^2+24\lambda_2\lambda_4-144g_2^2\lambda_4\right].
\qquad\label{betafunc_4}
\end{eqnarray}
Our results coincide with those obtained in Ref.~\cite{Hamada:2015bra} except for the $\lambda_3\lambda_4$-term in $\beta_{\lambda_3}$, which seems to be overlooked in that paper. But it does not affect perturbativity much. Additionally, we have rescaled the quartic couplings $\lambda_4$ by a factor $1/48$ and thus the numerical coefficients differ much from theirs. The beta functions in the SM are
\begin{eqnarray}
&&\beta_{g_1}^\mathrm{SM}=\frac{1}{16\pi^2}\frac{41}{10}g_1^3,~~
\beta_{g_2}^\mathrm{SM}=\frac{1}{16\pi^2}\left(-\frac{19}{6}\right)g_2^3,~~
\beta_{g_3}^\mathrm{SM}=\frac{1}{16\pi^2}(-7)g_3^3,
\label{SM:betafunc_1}\\
&&\beta_{y_t}^\mathrm{SM}=\frac{1}{16\pi^2}y_t\left(\frac{9}{2}y_t^2-\frac{9}{4}g_2^2-\frac{17}{20}g_1^2-8g_3^2\right),
\label{SM:betafunc_2}\\
&&\beta_{\lambda}^\mathrm{SM}=\frac{1}{16\pi^2}\left\{24\lambda^2-6y_t^4+\frac{3}{8}\left[2g_2^4+\left(g_2^2+\frac{3}{5}g_1^2\right)^2\right]+\lambda\left(-9g_2^2-\frac{9}{5}g_1^2+12y_t^2\right)\right\}.\nonumber\\*
\label{SM:betafunc_3}
\end{eqnarray}
From the weak scale to the septuplet threshold, couplings are evolving according to these functions.

Before heading towards the numerical study, we briefly introduce some numerical conventions. The $\overline{\mathrm{MS}}$ values of gauge couplings at the electroweak scale are given by~\cite{Beringer:1900zz}
\begin{eqnarray}
&&\alpha_s(m_Z)=\frac{g_s(m_Z)^2}{4\pi}=0.1184\pm0.0007,\\
&&\alpha(m_Z)=\frac{[g_2(m_Z)s_W(m_Z)]^2}{4\pi}=\frac{1}{127.926},\\
&&s_W^2=\sin^2\theta_W(m_Z)=0.2312.
\end{eqnarray}
The measured values of $y_t$ and $\lambda$ are actually determined from the observed masses of the top quark and the Higgs boson, $m_t$ and $m_h$, respectively.
Therefore, we need to derive the $\overline{\mathrm{MS}}$ values $y_t(m_Z)$ and $\lambda(m_Z)$  at the electroweak scale by the matching conditions~\cite{Hambye:1996wb}
\begin{equation}
y_t(\mu_0)=\frac{\sqrt{2}m_t}{v}[1+\delta_t(\mu_0)],~~
\lambda(\mu_0)=\frac{m_h^2}{2v^2}[1+\delta_h(\mu_0)],
\end{equation}
with setting $\mu_0 = m_Z$.
The related functions are
\begin{eqnarray}
\delta_t(\mu_0)&=&\left(-\frac{4\alpha_s}{4\pi}-\frac{4}{3}\frac{\alpha}{4\pi}+\frac{9}{4}\frac{m_t^2}{16\pi^2v^2}\right)\ln\frac{\mu_0^2}{m_t^2}+c_t,\\
\delta_h(\mu_0)&=&\frac{2v^2}{m_h^2}\frac{1}{32\pi^2v^4}[h_0(\mu_0)+m_h^2h_1(\mu_0)+m_h^4h_2(\mu_0)],\\
h_0(\mu_0)&=&-24m_t^4\ln\frac{\mu_0^2}{m_t^2}+6m_Z^4\ln\frac{\mu_0^2}{m_Z^2}+12m_W^4\ln\frac{\mu_0^2}{m_W^2}+c_0,\\
h_1(\mu_0)&=&12m_t^2\ln\frac{\mu_0^2}{m_t^2}-6m_Z^2\ln\frac{\mu_0^2}{m_Z^2}-12m_W^2\ln\frac{\mu_0^2}{m_W^2}+c_1,\\
h_2(\mu_0)&=&\frac{9}{2}\ln\frac{\mu_0^2}{m_h^2}+\frac{1}{2}\ln\frac{\mu_0^2}{m_Z^2}+\ln\frac{\mu_0^2}{m_W^2}+c_2.
\end{eqnarray}
The constants $c_0$, $c_1$, and $c_2$ are independent of the scale $\mu_0$, and their contributions to $\delta_h$ are less than 0.02. The constant $c_t$ lies in the range of $-0.052\leq c_t\leq-0.042$. Thus we neglect $c_0$, $c_1$ and $c_2$ and take $c_t=-0.052$ in the calculation. Choosing other values for $c_t$ would not essentially change our results.

\begin{figure}[!htbp]
\centering
\includegraphics[width=0.45\textwidth]{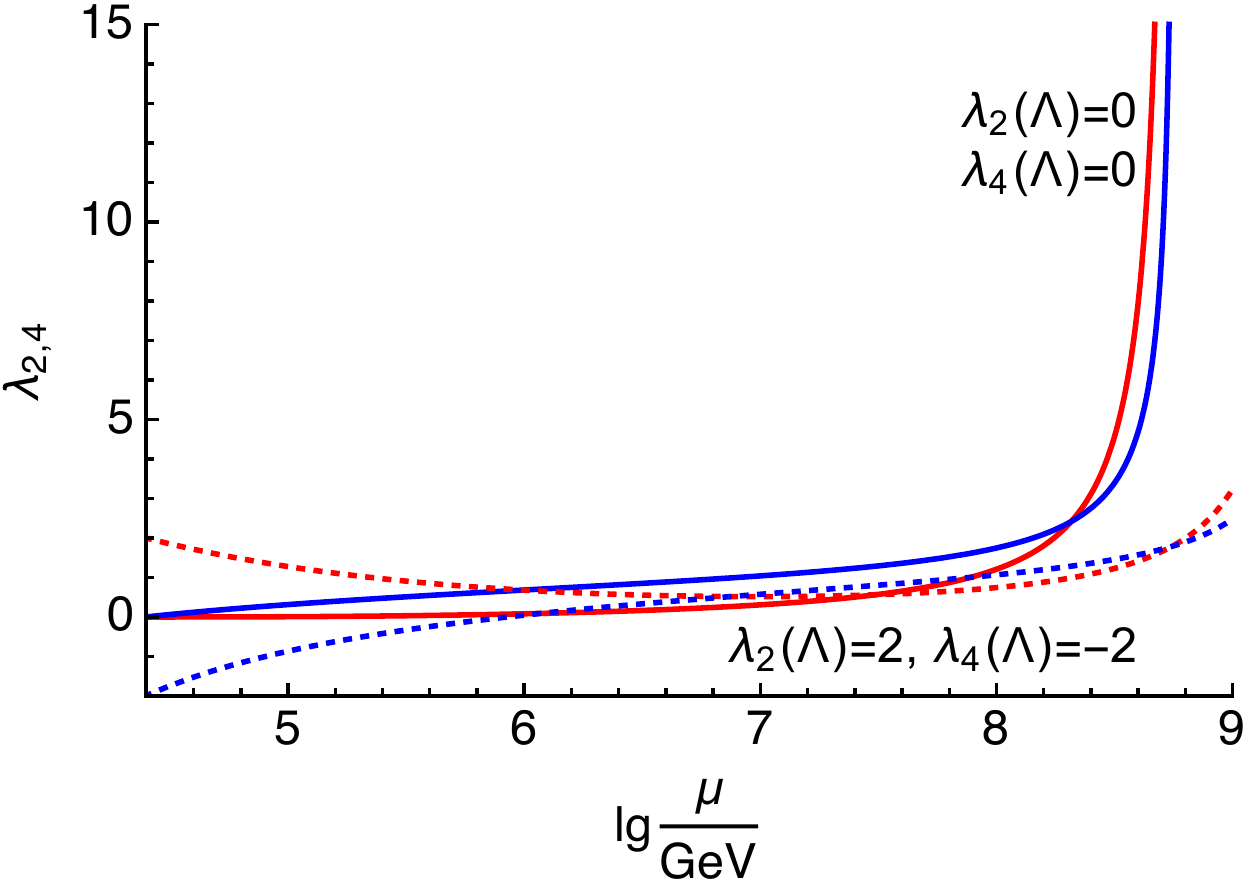}
\hspace{0.02\textwidth}
\includegraphics[width=0.45\textwidth]{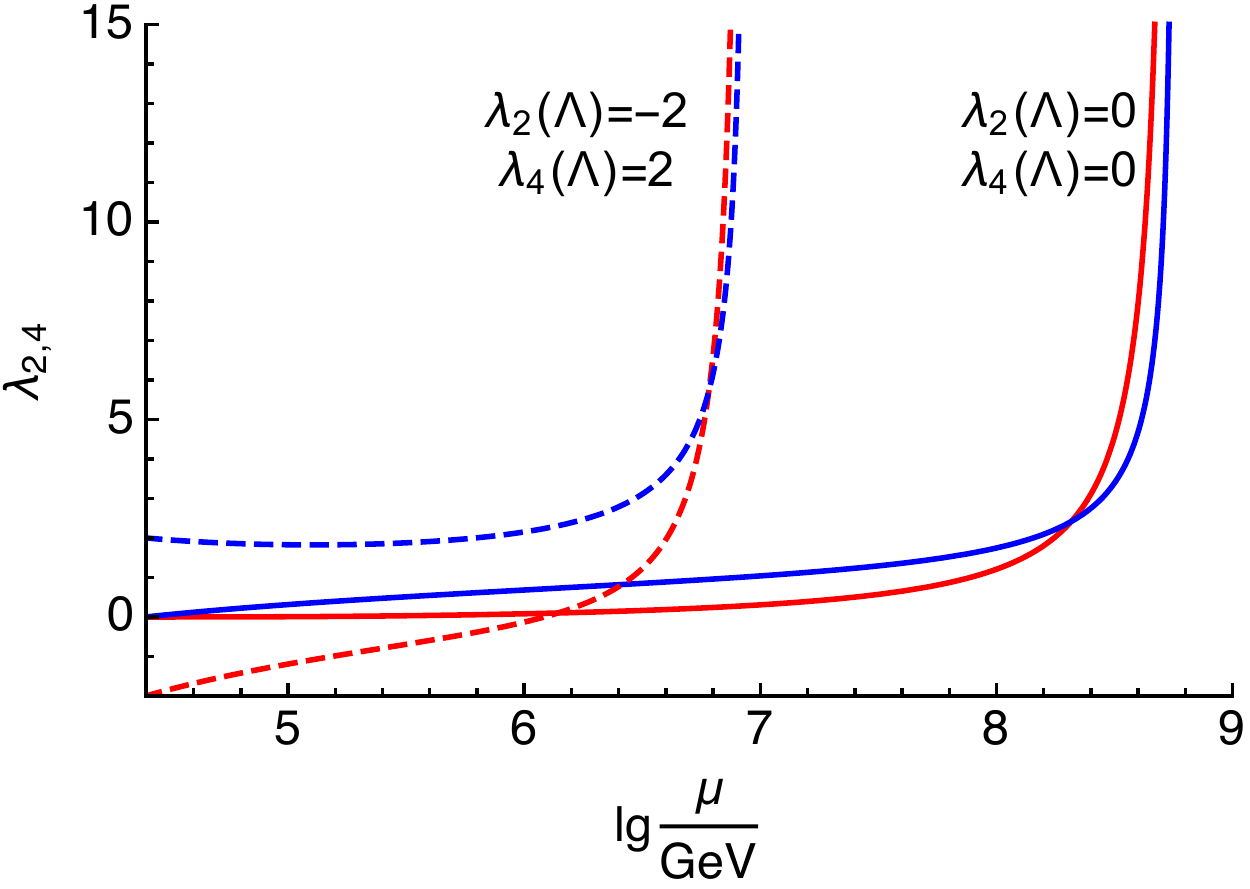}
\caption{The evolution of $\lambda_2$ (red lines) and $\lambda_4$ (blue lines) in the septuplet MDM model.}\label{running}
\end{figure}

It is illustrative to make some analysis of the RGEs~\eqref{betafunc_1}--\eqref{betafunc_4}. An analytical solution is impossible, despite an approximate solution treating $g_2$ as constant during running~\cite{Hamada:2015bra}~\footnote{The authors in Ref.~\cite{Hamada:2015bra} gave an estimation of the Landau pole as $\Lambda_{\rm LP}=1.0\times10^6\left(\frac{m}{100~\textrm{GeV}}\right)^{1.13}$.}. Even if the initial values of $\lambda_{2}$ and $\lambda_{4}$ are both zero, their Landau poles appear not very far from the septuplet threshold. The presence of the $\lambda_4$-term is crucial. The large positive contributions to $\beta_{\lambda_4}$ from the $g_2^4$- and $\lambda_4^2$-terms drive $\lambda_4$ increasing quickly. Then the terms involving $\lambda_4$ in $\beta_{\lambda_2}$ (with large positive coefficients) push $\lambda_2$ towards the Landau pole. As a matter of fact, $\lambda_2$ hits the Landau pole first. Given the septuplet MDM threshold $\Lambda= 8.8$~TeV or 25~TeV and initial values of $\lambda_2(\Lambda)=\lambda_4(\Lambda)=0$, we have $\Lambda_\mathrm{LP}\sim 10^{8}-10^{9}$~GeV. Here and henceforth, we set the septulet threshold $\Lambda$ to be $25$~TeV, and the evolution of $\lambda_2$ and $\lambda_4$ are indicated by the solid lines in Fig.~\ref{running}. This is consistent with the result in Ref.~\cite{Hamada:2015bra}. If the VS conditions were not imposed, it would be able to obtain a slightly higher $\Lambda_{\rm LP}\sim 10^{10}$ GeV by arranging a cancellation between the $\lambda_4\lambda_2$-term and $\lambda_2^2$-term in $\beta_{\lambda_2}$. For instance, This can be achieved if we set $\lambda_2(\Lambda)=2$ and $\lambda_4(\Lambda)=-2$, as demonstrated in the left panel of Fig.~\ref{running}.
On the other hand, if $\lambda_2(\Lambda)<0$ and $\lambda_4(\Lambda)>0$, the situation would be even worse, as shown in the right panel of Fig.~\ref{running}.

\subsection{Push up the Landau pole scale in the 7-3-5 model}

It is potential to push up the LP scale by introducing Yukawa couplings to the septuplet. This is inspired by the beta function structure of the Higgs self-coupling $\lambda$. It receives a large negative contribution from the top quark Yukawa coupling, which incurs the metastable problem of the electroweak vacuum at high scales. We will firstly present the 7-3-5 model, which is an extension to the septuplet model, and then analyze its implication to the Landau pole problem.

\subsubsection{The 7-3-5 model}

In order to construct Yukawa couplings to the septuplet, extra fermions should be introduced. Three minimal ways are available: $(1,0)\oplus(7,0)$, $(3,0)\oplus(5,0)$ and $(4,0)\oplus(4,0)$.
The first and second options have potential of explaining the tiny neutrino masses via the type-I and type-III seesaw mechanisms\footnote{There are some models with quintuplet fermion and septuplet scalar for generating neutrino masses at loop level~\cite{Cai:2011qr,Kumericki:2012bf,Culjak:2015qja}.}, respectively. However, we find that the first option is not as good as the second one due to a fairly large contribution to $\beta_{g_2}$ from the fermionic septuplet. On the other hand, in the last option, if the two $(4,0)$ fermions belong to the same chiral field (namely being Majorana fermions), it would lead to the problem of Witten global anomaly \cite{Witten:1982fp,Bar:2002sa}. Even worse, the Majorana mass term violates the accidental $Z_2$ symmetry of the septuplet and thus leads to the septuplet MDM radiatively decay into a pair of gauge bosons. A $(4,0)$ Dirac fermion avoids the Witten global anomaly, but it similarly endangers the stability of MDM at loop level. Therefore, in this work we concentrate on the $(3,0)\oplus(5,0)$ case, and the resulting model is dubbed as the 7-3-5 model.

In order to write down the Yukawa interaction terms, it is more convenient to use the tensor notation rather than the vector notation adopted above. More details about the tensor notation can be found in Ref.~\cite{Cai:2011qr}. The dictionary from the latter to the former notation reads
\begin{eqnarray}
&&\Phi=\frac{1}{\sqrt{2}}\begin{pmatrix}\Delta^{(3)}\\ \Delta^{(2)}\\ \Delta^{(1)}\\ \Delta^{(0)}\\ \Delta^{(-1)}\\ \Delta^{(-2)}\\ \Delta^{(-3)}\end{pmatrix}=\frac{1}{\sqrt{2}}\begin{pmatrix}\Phi_{111111}\\ \sqrt{6}\Phi_{111112}\\ \sqrt{15}\Phi_{111122}\\ \sqrt{20}\Phi_{111222}\\ -\sqrt{15}\Phi_{112222}\\ \sqrt{6}\Phi_{122222}\\ -\Phi_{222222}\end{pmatrix},\\
&&\Psi_{L}=\begin{pmatrix}\Psi_{+2,L}\\ \Psi_{+1,L}\\ \Psi_{0,L}\\ \Psi_{-1,L}\\ \Psi_{-2,L}\end{pmatrix}=\begin{pmatrix}\Psi_{1111,L}\\ 2\Psi_{1112,L}\\ \sqrt{6}\Psi_{1122,L}\\ -2\Phi_{1222,L}\\ \Phi_{2222,L}\end{pmatrix},~~
\Sigma_{R}=\begin{pmatrix}\Sigma_{+1,R}\\ \Sigma_{0,R}\\ \Sigma_{-1,R}\end{pmatrix}=\begin{pmatrix}\Sigma_{11,R}\\ \sqrt{2}\Sigma_{12,R}\\ -\Sigma_{22,R}\end{pmatrix}.
\end{eqnarray}
Here we have assigned the left and right chiralities to the quintuplet and triplet, respectively. At renormalizable level the most generic Yukawa interactions can be written down as
\begin{eqnarray}\label{yukawa}
\mathcal{L}_\mathrm{yuk}=&-\sqrt{15}y\Phi_{ijklmn}\overline{\Psi_L^{ijkl}}\Sigma_{R,m'n'}\varepsilon^{mm'}\varepsilon^{nn'}\cr
&-(y_\Sigma)_{ab} \overline{l_{a,L}^i} (\Sigma_{b,R})_{ij} H_k\varepsilon^{jk}+ \mathrm{h.c.},
\end{eqnarray}
where $i,j,k,l,m,n=1,2$ are $SU(2)_L$ tensor indices and  symmetric for $\Phi$, $\Psi$, and $\Sigma$. $a$ and $b$ are family indices and for the sake of realistic neutrino mixing at least two triplets are required. Terms in the second line, along with Majorana mass terms for the fermions, constitute the type-III seesaw mechanism. The Yukawa couplings $y_\Sigma$ are irrelevantly small. Note that the 7-dimensional representation from the decomposition $5\times5=1_S+3_A+5_S+7_A+9_S$ is antisymmetric, thus the coupling $\Phi\Psi\Psi$ vanishes. In the vector notation, Eq.~\eqref{yukawa} gives the Yukawa couplings of the septuplet as
\begin{eqnarray}
\mathcal{L}_\mathrm{yuk}&=&\frac{y}{\sqrt{2}}\{[\Delta^{(0)}\sqrt{3}\overline{\Psi_{-1}}\Sigma_{-1}+\Delta^{(-1)}(-\sqrt{6}\overline{\Psi_0}\Sigma_{+1}-2\sqrt{2}\overline{\Psi_{-1}}\Sigma_0+\overline{\Psi_{-2}}\Sigma_{-1})\nonumber\\
&&\qquad~ +\Delta^{(-2)}(-\sqrt{10}\overline{\Psi_{-1}}\Sigma_{+1}-\sqrt{5}\overline{\Psi_{-2}}\Sigma_0)-\sqrt{15}\Delta^{(-3)}\overline{\Psi_{-2}}\Sigma_{+1}+\mathrm{h.c.}]
\nonumber\\
&&\qquad -3\Delta^{(0)}\overline{\Psi_0}\Sigma_0\}.
\end{eqnarray}
Where $\Psi_{-Q}=\Psi_{-Q,L}+(\Psi_{+Q,L})^c$ and $\Sigma_{-Q}=\Sigma_{-Q,L}+(\Sigma_{+Q,L})^c$.
The gauge couplings of the fermions are
\begin{eqnarray}
\mathcal{L}_\mathrm{gauge}&=&g_2[(\sqrt{3}\overline{\Psi_0}\gamma^\mu\Psi_{-1}+\sqrt{2}\overline{\Psi_{-1}}\gamma^\mu\Psi_{-2})W^+_\mu+\overline{\Sigma_{0}}\gamma^\mu\Sigma_{-1}W^+_\mu+\mathrm{h.c.}]\nonumber\\
&&+g_2(\sum_{Q=1}^2Q\overline{\Psi_Q}\gamma^\mu\Psi_{Q}+\overline{\Sigma_{+1}}\gamma^\mu\Sigma_{+1})W^3_\mu.
\end{eqnarray}

The presence of the quintuplet and triplet further modifies the one-loop beta functions with the following extra terms:
\begin{eqnarray}
\delta \beta_{g_2}&=&\frac{8g_2^3}{16\pi^2},\label{betag2:735}\\
\delta \beta_{\lambda_2}&=&\frac{1}{16\pi^2}(-54y^4+40y^2\lambda_2),\\
\delta \beta_{\lambda_3}&=&\frac{20y^2\lambda_3}{16\pi^2},\\
\delta \beta_{\lambda_4}&=&\frac{1}{16\pi^2}(-96y^4+40y^2\lambda_4),\\
\beta_y &=&\frac{y}{16\pi^2}(25y^2-24g_2^2)\label{betay}.
\end{eqnarray}
As expected, the Yukawa coupling $y$ has negative contributions to $\beta_{\lambda_2}$ and $\beta_{\lambda_4}$. Below we study how the LP scale can be substantially pushed up by this coupling.

\subsubsection{How high can the Landau pole scale be?}

Now we have two free parameters at hand. One is $y$; the other one is $M_{35}$, the threshold
above which the 7-3-5 Yukawa coupling becomes active. The mass scales of the quintuplet and triplet are set to be equal to $M_{35}$~\footnote{In practice, $M_{35}$ can be identified as the heavier one. But the situation is actually worse since the lighter fermion just induces increase in $g_2$ from its mass to $M_{35}$.}. In order to push up $\Lambda_{\rm LP}$ as high as possible, we find that $M_{35}$ should not be far above the MDM mass scale, and moreover the initial value of $y(M_{35})$ should be fine-tuned. Thereby, the 7-3-5 model does not provide a quite satisfactory alleviation to the Landau pole problem.

First of all, the Landau pole of $g_2$ is of concern. As shown in Eq.~\eqref{betag2:735}, the beta function of $g_2$ receives large positive contributions from the quintuplet and triplet. Consequently, a low $M_{35}$ would drive $g_2$ to the Landau pole quickly~\footnote{To maximize the Landau pole scale of $g_2$, we assume a hierarchy between different triplet generations and only the lightest one is active at the low energy scales we concern.}.
The explicit solution is
\begin{eqnarray}
&&\alpha_2^{-1}(\mu)=
\alpha_2^{-1}(m_Z)-\frac{b_2^\mathrm{SM}}{2\pi}\ln\frac{\Lambda}{m_Z}
-\frac{b_2^\mathrm{sep}}{2\pi}\ln\frac{M_{35}}{\Lambda}-\frac{b_2^\mathrm{tot}}{2\pi}\ln\frac{\mu}{M_{35}},
\end{eqnarray}
where $b_2^\mathrm{SM}=-19/6$, $b_2^\mathrm{sep}=3/2$, and $b_2^\mathrm{tot}=19/2$. From it one can determine the Landau pole scale of $g_2$:
\begin{eqnarray}
\Lambda^{(g_2)}_{\rm LP}=M_{35}\left(\frac{\Lambda}{M_{35}}\right)^{b_2^\mathrm{sep}/b_2^\mathrm{tot}}\exp\left[\frac{2\pi}{b_2^\mathrm{tot}}\alpha_2^{-1}(\Lambda)\right].
\end{eqnarray}
For example, if $M_{35}=10^5$~GeV, we have $\Lambda^{(g_2)}_{\rm LP}=1.54\times10^{14}$~GeV, far below the Planck scale. Note that typically the non-perturbative scale is only slightly lower than $\Lambda_{\rm LP}$, so we do not distinguish them in this paper.

One cannot rely on increasing $M_{35}$ to lift $\Lambda^{(g_2)}_{\rm LP}$. As our purpose is to push up the Landau pole scales of $\lambda_2$ and $\lambda_4$, $M_{35}$ is forced to be not far from the septuplet threshold $\Lambda$. Otherwise, $\lambda_2$, perhaps as well as $\lambda_4$, would quickly evolve to a sufficiently large value such that $y(M_{35})\gg 1$ is needed to slow down the running of $\lambda_2$ and $\lambda_4$. Moreover, $y$ itself grows very fast and eventually diverges at $\Lambda^{(y)}_{\rm LP}$, which could be much lower than $\Lambda^{(g_2)}_{\rm LP}$. We can see this from the explicit expression of $y(t)$, where $t\equiv\ln(\mu/M_{35})$ denotes the logarithm of the energy scale. In practice, RGE~\eqref{betay} can be analytically solved:
\begin{align}\label{LP:y}
y^2(t)=\frac{(24+b_2)g^2_2(0)}{F_0 [g_2(t)/g_2(0)]^{48/b_2}
+25 [g_2(t)/g_2(0)]^{-2}},
\end{align}
with $F_0\equiv (24+b_2) g^2_2(0)/y^2(0)-25 $ and $b_2=b_2^\mathrm{tot}=19/2$.

\begin{figure}[htb]
\centering
\includegraphics[width=0.6\textwidth]{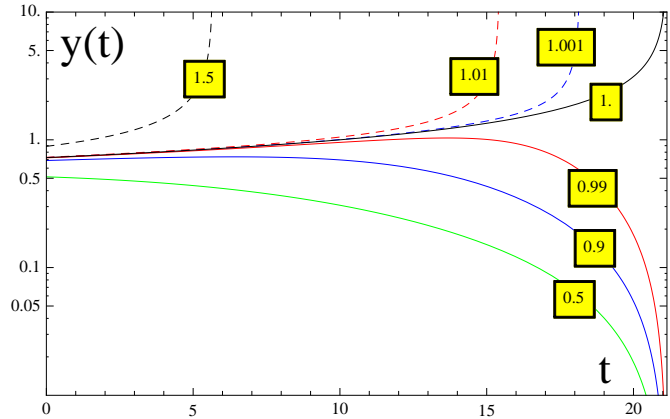}
\caption{Running of $y(t)$ for different choices of $y(0)$, which is the labels on the lines in unit of $y_c$. The LP scale of $g_2$ is at the maximum of the $t$-axis. We have chosen $M_{35}=10^{5.5}$ GeV, but the curves are not sensitive to this value.}\label{y:run}
\end{figure}

As long as we have a large $y(0)$ such that $F_0<0$, $y(t)$ will meet the Landau pole as $g_2(t)$ evolves to the value
\begin{align}\label{LP:y2}
g_2^2(t_\mathrm{LP}^{(y)})\simeq g^2_2(0)\left( \frac{25}{|F_0|}\right)^{b_2/(24+b_2)}= &g^2_2(0) \left[1-\frac{(24+b_2)g^2_2(0)}{25y^2(0)}\right]^{-b_2/(24+b_2)}.
\end{align}
Based on this equation, it is straightforward to derive an expression for $\Lambda^{(y)}_{\rm LP}$:
\begin{eqnarray}
\Lambda_{\rm LP}^{(y)}=\frac{\Lambda_{\rm LP}^{(g_2)}}{\exp\left[\dfrac{8\pi^2}{b_2g_2^2(0)}\left(\dfrac{-F_0}{25}\right)^{b_2/(24+b_2)}\right]}.
\end{eqnarray}
Critical values that lead to $F_0=0$ and $\Lambda_\mathrm{LP}^{(y)}=\Lambda_\mathrm{LP}^{(g_2)}$ are $\pm y_c$, where
\begin{align}\label{yc}
y_c\equiv \left(\frac{24+b_2}{25}\right)^{1/2}g_2(0).
\end{align}
Thus $F_0=25[y_c^2/y^2(0)-1]$.
Since $\Lambda_{\rm LP}^{(y)}$ is exponentially suppressed compared to $\Lambda_{\rm LP}^{(g_2)}$, a small derivation of $F_0$ from zero will lead to a dramatically smaller $\Lambda_{\rm LP}^{(y)}$, as shown in Fig.~\ref{y:run}, where the labels on the lines are values of $y(0)$ in unit of $y_c$ and $y(0)>y_c$ corresponds to $F_0<0$.
In other words, in order to retain a high $\Lambda_{\rm LP}^{(y)}$, $y(0)$ must be extremely close to $\pm y_c$.
This raises a fine-tuning problem.

\begin{figure}[htb]
\centering
\includegraphics[width=1.0\textwidth]{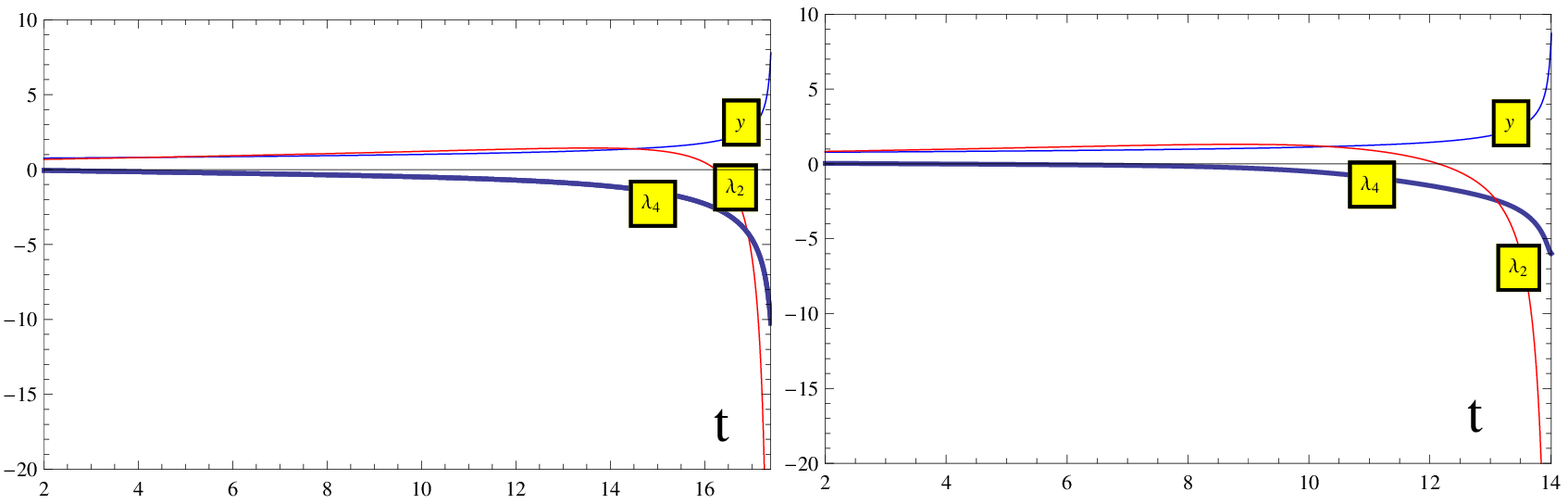}
\includegraphics[width=1.0\textwidth]{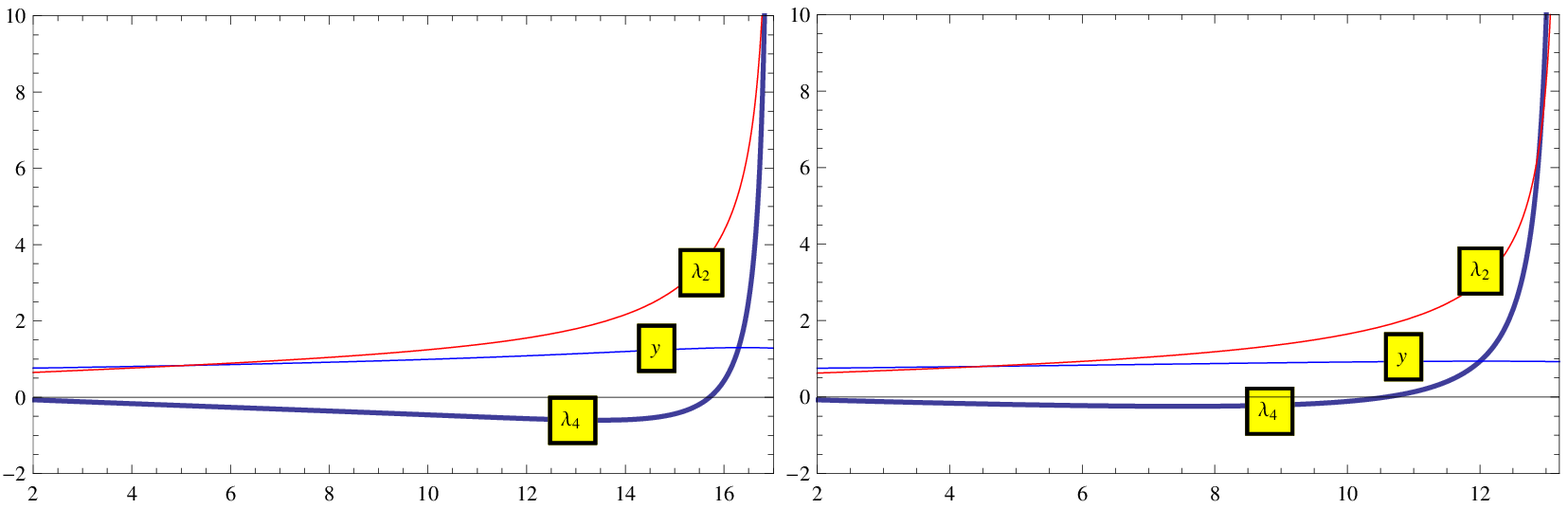}
\caption{RGE running of couplings in four cases. Top left: $y(0)=1.001 y_c$ with $M_{35}=10^{5.7}$~GeV; Top right: $y(0)=1.01 y_c$ with $M_{35}=10^{6.1}$~GeV; Bottom left: $y(0)=0.999 y_c$ with $M_{35}=10^{5.6}$~GeV; Bottom right: $y(0)=0.99 y_c$ with $M_{35}=10^{5.5}$~GeV. }\label{run:y0}
\end{figure}

Unfortunately, this kind of fine-tuning is robust in any case.
In the case of $F_0>0$, $y(t)$ no longer has a Landau pole but instead become asymptotic free as $g_2(t)$ approaches its Landau pole. Due to the large power $48/b_2\approx 5.1$, $y(t)$ will quickly run to zero once the first term of the denominator in Eq.~(\ref{LP:y}) dominates. We demonstrate this behavior for several $y(0)$ values in Fig.~\ref{y:run}, where $y(0)<y_c$ corresponds to $F_0>0$. One can clearly see that $y(0)$ must be also sufficiently close to the critical value $y_c$, otherwise the Yukawa effect becomes negligible soon and the LP scales of $\lambda_2$ and $\lambda_4$ cannot be pushed up.
Therefore, this case is not supposed to be better than the previous case. But it has an advantage that $\lambda_2$ and $\lambda_4$ are positive near their Landau poles. By contrast, in the $F_0<0$ case they run to large negative values near the Landau pole of $y(t)$. We can see these features in Fig.~\ref{run:y0}. We leave an analytical understanding of the Landau poles of $\lambda_2$ and $\lambda_4$ in the next subsection.

From this analysis we can learn how to choose values of $M_{35}$ that can lead to the highest LP scales. It should be chosen to guarantee a negative $\beta_{\lambda_{2}}(0)\sim -\mathcal{O}(0.01)$ such that the most intractable coupling $\lambda_{2}$ is tamed. Bear in mind that since $y(0)$ has been almost fixed around 0.7 by the condition $F_0\approx 0$, there is a strong upper bound on $M_{35}$. In terms of our numerical investigation, the highest $M_{35}$ is larger than the septuplet threshold $\Lambda$ by about (typically slightly less than) 2 orders of magnitude. For instance, for $m_{\rm DM}=22$ TeV, it is found that $M_{35}\approx 10^{5.7}$~GeV is required and the resulting maximal Landau pole scale is $\Lambda_{\rm LP}\approx 10^{13}$ GeV if we tolerate $y(0)=1.001 y_c$;  If we only tolerate $y(0)=1.01 y_c$, the maximal Landau pole is $\Lambda_{\rm LP}\approx 10^{12}$ GeV with $M_{35}\approx 10^{6.1}$ GeV. We show the running in these two cases in Fig~\ref{run:y0}. For comparison, we also plot two cases with $F_0>0$: $y(0)=0.99 y_c$ with $M_{35}=10^{5.5}$ GeV and $\Lambda_{\rm LP}\approx 1.7\times 10^{11}$ GeV; $y(0)=0.999 y_c$ with $M_{35}=10^{5.6}$ GeV and $\Lambda_{\rm LP}\approx 9.6\times 10^{11}$ GeV.

\subsubsection{Constraints from perturbativity and VS in the $F_0=0$ case}

As a demonstration of the impacts of perturbativity and as well VS on the model, here we assume the ideal limit for $F_0$, i.e., it is exactly zero and then one can obtain the highest LP scale. For the given septuplet MDM mass 25~TeV, it means that we have to fine-tune the initial $y(0)=0.7286422$.

\begin{figure}[!htbp]
\centering
\includegraphics[width=0.45\textwidth]{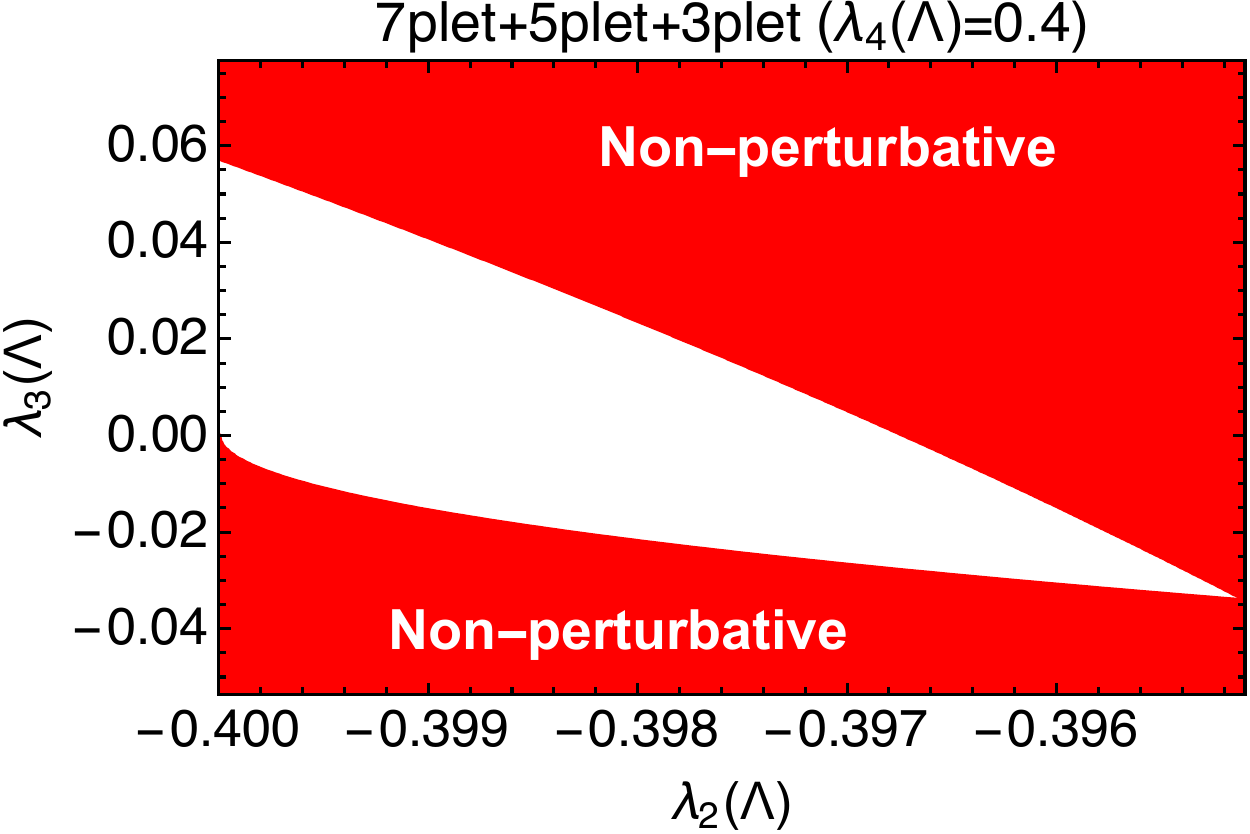}
\hspace{0.02\textwidth}
\includegraphics[width=0.43\textwidth]{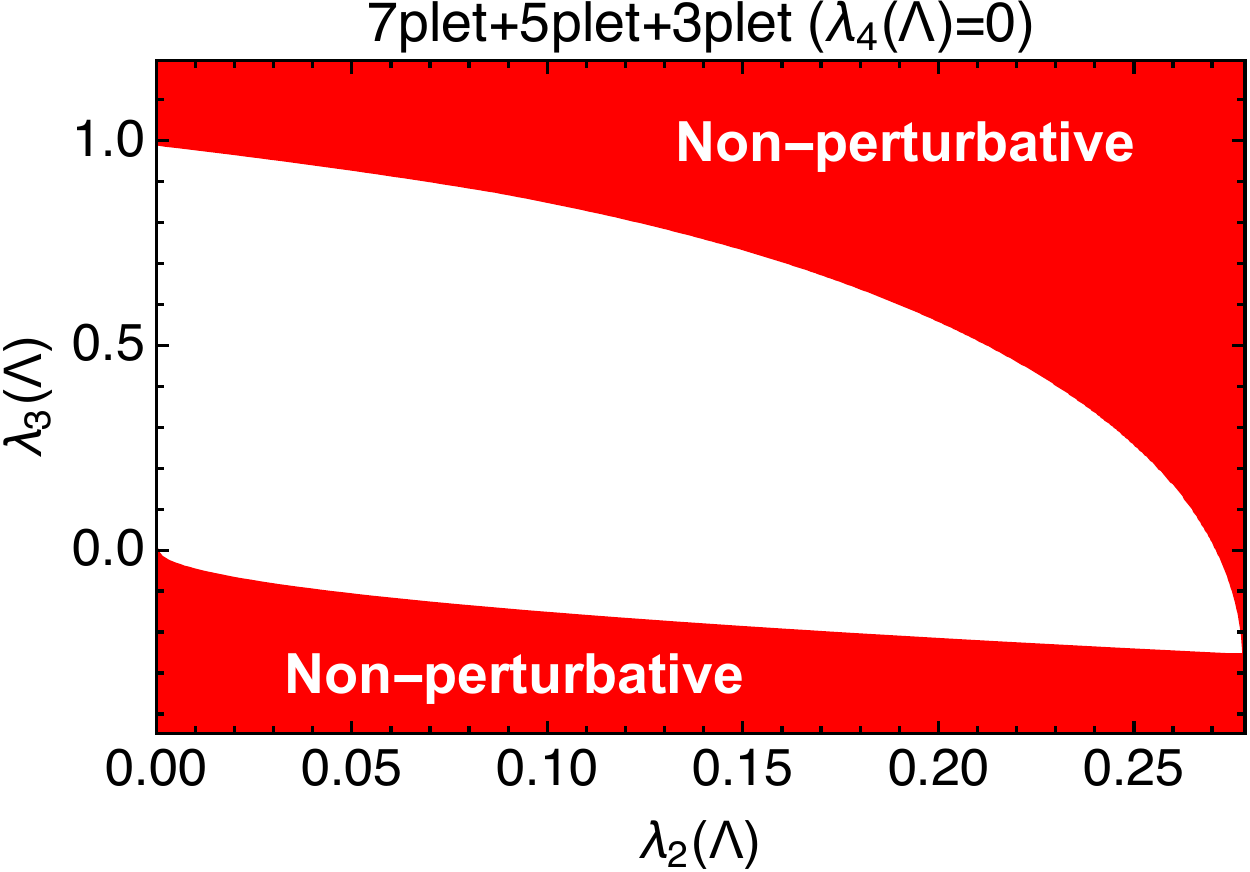}\\
\hspace{0.02\textwidth}
\includegraphics[width=0.48\textwidth]{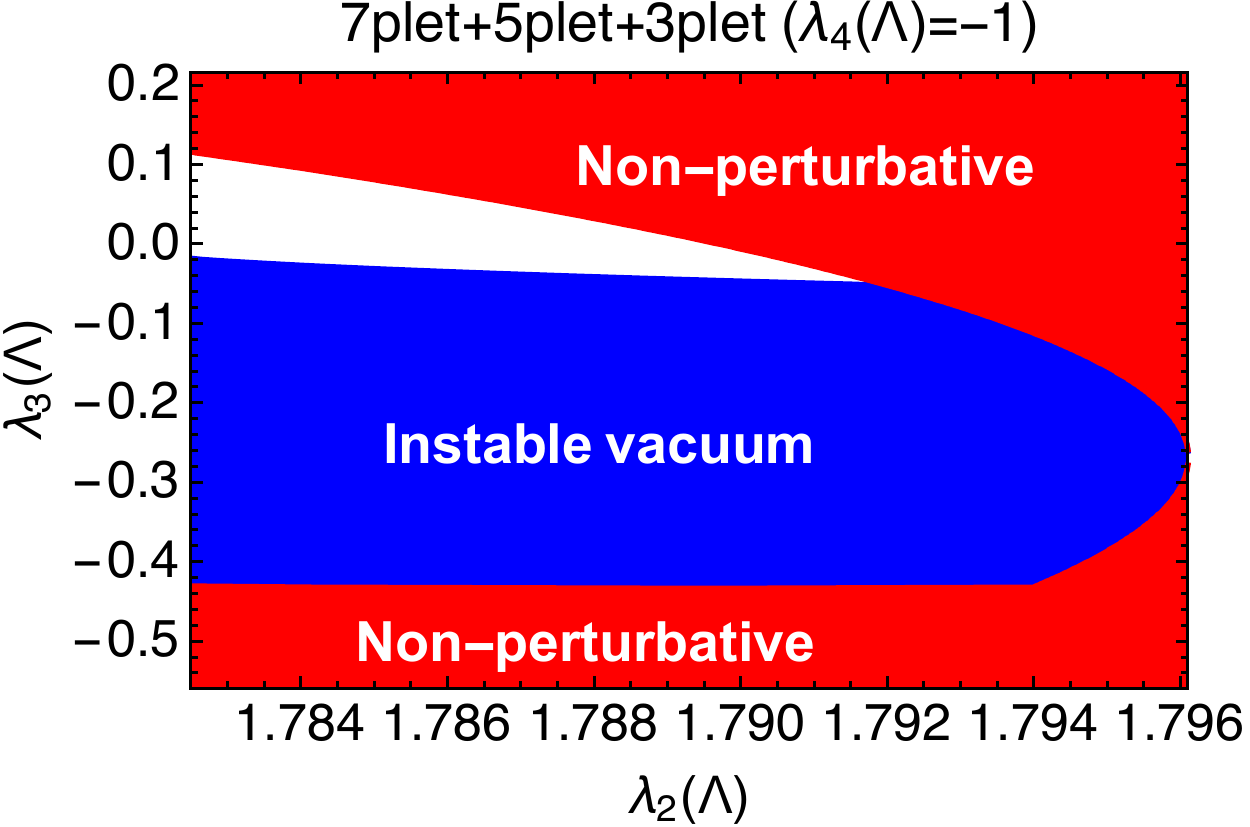}
\caption{Regions excluded by the VS (blue) and perturbativity (red) conditions in the $\lambda_2(\Lambda)$-$\lambda_3(\Lambda)$ plane for the 7-3-5 model.
The top-left (top-right) panel corresponds to $\lambda_4(\Lambda)=0.4~(0)$, while the bottom panel corresponds to $\lambda_4(\Lambda)=-1$.}\label{case2}
\end{figure}

We survey 3 slices of the parameter space corresponding to $\lambda_4(\Lambda)=0.4$, $0$, $-1$ as input, and evolve the couplings to a cutoff scale of $10^{14}$~GeV, just a little bit below the highest LP scale, and then impose the VS and perturbativity conditions to give constraints. Here the perturbativity conditions mean that the absolute value of any coupling cannot exceed $4\pi$. We assume $M_{35}=10^{5.5}$~GeV and the results are shown in Fig.~\ref{case2}, where the blue (red) regions are excluded by the VS (perturbativity) conditions, while the white regions can fulfill both the conditions. These plots have the following features.

\begin{itemize}
\item When $\lambda_4(\Lambda)=0.4$, the acceptable range of $\lambda_3(\Lambda)$ that fulfill the two conditions is $-0.034\lesssim \lambda_3(\Lambda)\lesssim 0.057$. On the other hand, $\lambda_2(\Lambda)$ is bounded as $-0.4\leq\lambda_2(\Lambda)\lesssim -0.395$, which is very narrow. $0.4$ is almost the upper bound on $\lambda_4(\Lambda)$. If $\lambda_4(\Lambda)$ has a larger value, it will grow too fast and cannot keep perturbative up to the cutoff scale.
\item When $\lambda_4(\Lambda)=0$, the region survived in the $\lambda_2(\Lambda)$-$\lambda_3(\Lambda)$ plane is maximized. The acceptable range of $\lambda_3(\Lambda)$ is $-0.247\lesssim \lambda_3(\Lambda)\lesssim 0.995$, which is denoted by the green band in the $m_0$-$\lambda_3$ plane in Fig.~\ref{m0l3}. On the other hand, the acceptable range of $\lambda_2(\Lambda)$ is enlarged as $0\leq\lambda_2(\Lambda)\leq 0.278$. Notice that the whole perturbative region satisfies the VS conditions when $\lambda_4(\Lambda)\geq0$.
\item When $\lambda_4(\Lambda)=-1$, the range of $-0.047< \lambda_3(\Lambda)< 0.116$ is acceptable, and $\lambda_2(\Lambda)$ is bounded by the two conditions as $1.783\leq\lambda_2(\Lambda)\leq 1.792$, which are the maximal values that $\lambda_2(\Lambda)$ can be in the 7-3-5 model. If $\lambda_4(\Lambda)$ becomes smaller, the vacuum instability region will enlarge and leave no more acceptable region. Therefore, $-1$ is basically the lower bound on $\lambda_4(\Lambda)$.
\end{itemize}

In order to further understand these results, we present an analytical analysis in \S\ref{subsectF0=0}. The existence of perturbative and vacuum stable parameter regions relies on the fact that the solutions to quartic coupling RGEs could finally converge to a simple pattern in high energy scales. They are all asymptotically proportional to $g_2^2(\mu)$ and fortunately these asymptotic solutions trivially satisfy the VS conditions~\eqref{VS1} or \eqref{VS2}. Note that for $\lambda_4(\Lambda)<0$ there are some regions satisfying the perturbativity conditions but excluded by the VS conditions. Even so, this is not conflict with the above statement,
because these vacuum instability regions appear before the quartic couplings converge to their asymptotic solutions.

\subsubsection{Towards the Planck scale: relaxing the septuplet mass}

One of the main merits of the MDM model is that it predicts a unique mass for thermally produced DM particles via the observed relic abundance. But we may give it up and turn to other production mechanisms rather than the conventional freeze-out mechanism, e.g., freeze-in much lighter MDM~\cite{Aoki:2015nza}. Then, the MDM particle mass can be relaxed. For pushing up the LP scale, a much heavier MDM particle is favored. One may worry about the correct MDM relic density because heavier MDM could not quickly annihilate and thus would overclose the Universe. However, if MDM is very heavy, for instance, at PeV or even higher scales, it is reasonable to conjecture that MDM is not abundantly produced, provided that the reheating temperature is even lower than the mass scale of the MDM particle. Thus one may utilize some nonthermal ways to produce the correct relic abundance. A detailed study is beyond the scope of this paper, and we refer to some relevant studies~\cite{Chung:1998zb,Falkowski:2012fb}.

Let us consider an example with $m_0=10^8$~GeV. We find that $M_{35}$ should be chosen lower than $10^{9.5}$~GeV to prevent disastrous growing behaviors of the quartic couplings. Now the LP scale can be higher than $10^{15}$~GeV and we should take into account the second generation triplet. Thus the running of $g_2$ obtains an extra correction and becomes faster. The highest LP scale, which is actually $\Lambda_\mathrm{LP}^{(g_2)}$, can be determined by
\begin{equation}
0=\alpha_2^{-1}(\Lambda_\mathrm{LP}^{(g_2)}) =
\alpha_2^{-1}(m_Z)-\frac{b_2^\mathrm{SM}}{2\pi}\ln\frac{\Lambda}{m_Z} -\frac{b_2^\mathrm{sep}}{2\pi}\ln\frac{\Lambda_1}{\Lambda} -\frac{b_2^{(1)}}{2\pi}\ln\frac{\Lambda_2}{\Lambda_1} -\frac{b_2^\mathrm{(2)}}{2\pi}\ln\frac{\Lambda_\mathrm{LP}^{(g_2)}}{\Lambda_1},
\end{equation}
where $b_2^{(1)}=19/2$ and $b_2^{(2)}=65/6$. $\Lambda_1$ ($\Lambda_2$) corresponds to the threshold scale of the first (second) generation triplet. The numerical result is $\Lambda_\mathrm{LP}^{(g_2)}\simeq 10^{21}$~GeV, which is higher than the Planck scale. Then even in the ${F_0}\neq0$ case for $y_{(2)}(\Lambda_2)$ (the Yukawa coupling for the second generation triplet), the model is still possible to remain perturbative up to the Planck scale.

\subsection{Analytical treatments}
\subsubsection{$F_0=0$}\label{subsectF0=0}

In the special case of $F_0=0$, the evolution of $y(t)$ exactly follows that of $g_2(t)$, and we can reach the maximal LP scale $\Lambda^{(g_2)}_{\rm LP}$. Although there is no particular theoretical motivation, it would be illustrative to investigate such an ideal case.

If ${F_0}=0$, from Eq.~(\ref{LP:y}) one gets a simple solution $y^2(t)=(24+b_2)g_2^2(t)/25$. Substitute $y^2(t)$ into the beta functions of ${\lambda_i}$ and $\lambda$ and neglect the subdominant contributions from $y_t$ and $g_1$, we obtain the following RGEs:
\begin{eqnarray}
\frac{d\lambda_2}{dt'}&=&30\lambda_2^2+\frac{45}{2}\lambda_4^2+51\lambda_2\lambda_4-\frac{452}{5}g_2^2\lambda_2-\frac{121203}{1250}g_2^4+2\lambda_3^2, \label{blam2}\\
\frac{d\lambda_4}{dt'}&=&\frac{72264}{625}g_2^4-\frac{452}{5}g_2^2\lambda_4+\frac{255}{8}\lambda_4^2+24\lambda_2\lambda_4, \label{blam4}\\
\frac{d\lambda_3}{dt'}&=&36g_2^4+\lambda_3\left[-\frac{497}{10}g_2^2+18\lambda_2+4\lambda_3+\frac{51}{2}\lambda_4+12\lambda\right], \label{blam3}\\
\frac{d\lambda}{dt'}&=&24\lambda^2+\frac{9}{8}g_2^4-9\lambda g_2^2+\frac{7}{2}\lambda_3^2. \label{blam}
\end{eqnarray}
We have rescaled $t'=t/16\pi^2$ to drop the annoying factor $16\pi^2$ for simplicity.

It is interesting to notice that these differential equations admit a particular solution where all couplings follow the running of $g_2^2(t')$~\footnote{Here the key points are $d g_2/dt'\propto g_2^3$ and the RGEs essentially only involve scalar quartic couplings. In this sense, such solutions are generic for the scalar system with gauge interactions.}:
\begin{eqnarray}\label{assumption1}
\lambda_2(t')=a_1g_2^2(t'),~\lambda_4(t')=a_2g_2^2(t'),~\lambda_3(t')=a_3g_2^2(t'),~\lambda(t')=a_4g_2^2(t'),
\end{eqnarray}
where all $a_i$ are constant. Substitute them into Eqs.~\eqref{blam2}--\eqref{blam}, and these differential equations become algebraic equations:
\begin{eqnarray}
2b_2a_1&=&30a_1^2+\frac{45}{2}a_2^2+51a_1a_2-\frac{452}{5}a_1-\frac{121203}{1250}+2a_3^2, \label{algeq1}\\
2b_2a_2&=&\frac{72264}{625}-\frac{452}{5}a_2+\frac{255}{8}a_2^2+24a_1a_2, \label{algeq2}\\
2b_2a_3&=&36+a_3[-\frac{497}{10}+18a_1+\frac{51}{2}a_2+4a_3+12a_4], \label{algeq3}\\
2b_2a_4&=&24a_4^2+\frac{9}{8}-9a_4+\frac{7}{2}a_3^2. \label{algeq4}
\end{eqnarray}
There are two sets of solutions,
\begin{eqnarray}
(a_1,a_2,a_3,a_4)&=&(-0.826831,1.33252,1.05752,0.944318),\\
 (a_1,a_2,a_3,a_4)&=& (-0.839066,1.32379,0.795673,0.134918).
\end{eqnarray}
Both of them satisfy the VS conditions. The second solution, which gives a small coefficient of $\lambda$, is closer to the exact numerical result and thus will be used.

\begin{figure}[!htbp]
\centering
\includegraphics[width=1\textwidth]{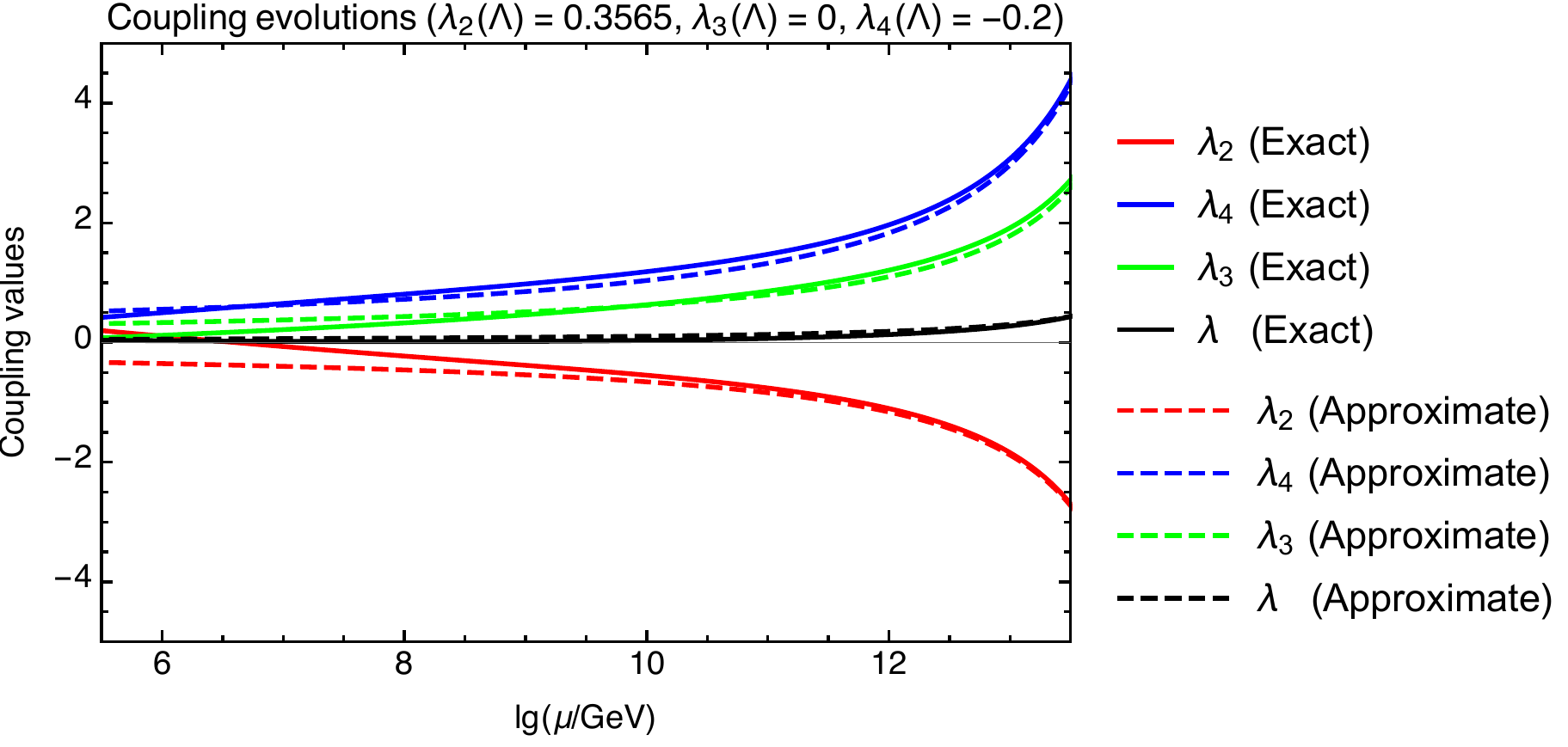}
\caption{Evolution of $\lambda_2$, $\lambda_3$, $\lambda_4$, and $\lambda$ in the 7-3-5 model for $M_{35}=10^{5.5}$~GeV. The solid lines are the exact numerical solutions, while the dashed lines represent the approximate analytical solutions based on the assumption~\eqref{assumption1}.}\label{runningcoupling}
\end{figure}

In the above treatment we do not refer to initial conditions, so it is not a surprise that they are only suitable for describing the evolution of couplings at sufficiently high energy scales, where the influence of initial conditions has been almost erased. This is closely related to the fact that $g_2(t')$ steadily increases towards high energy scales after the septuplet is involved to make $b_2>0$. To visualize this behavior, we demonstrate a comparison between the solutions from exact numerical calculation and from the approximate analytical solutions in Fig.~\ref{runningcoupling}. We can see that although some parameters are far away from the analytical solutions at the beginning, they all converge to these solutions at high energy scales.

\subsubsection{$F_0>0$}

When ${F_0}>0$, the Landau pole of $y$ no longer exists. Instead, $y$ becomes a non-monotonic function of the energy scale $\mu$. It increases in the interval
\begin{eqnarray}
M_{35}\leq\mu<M_{35}\exp\left\{\frac{8\pi^2}{b_2g_2^2(0)}\left[1-\left(\frac{24F_0}{25b_2}\right)^{b_2/(24+b_2)}\right]\right\}
\end{eqnarray}
and decreases in the interval
\begin{eqnarray}
M_{35}\exp\left\{\frac{8\pi^2}{b_2g_2^2(0)}\left[1-\left(\frac{24F_0}{25b_2}\right)^{\frac{b_2}{24+b_2}}\right]\right\}\leq\mu<\Lambda_\mathrm{LP}^{(g_2)}.
\end{eqnarray}
In the latter interval, $y^2$ behaves as $$\frac{24+b_2}{{F_0}}\left(\frac{g_2(t')}{g_2(t'_0)}\right)^{-48/b_2}g_2^2(t'_0)$$ when $g_2^2\gg g_2^2(t'_0)$. It decrease very quickly and finally goes to zero. Then the 7-3-5 model essentially turns back to the SM+septuplet model at high energy scales. As we known, $\lambda_2$ and $\lambda_4$ will grow faster than $g_2^2$ and reach their Landau poles before $\Lambda_\mathrm{LP}^{(g_2)}$. To illustrate this more concretely, we will solve the RGEs at high energy scales. The equations are
\begin{eqnarray}
\frac{d(g_2^2)}{dt'} &=& 2b_2g_2^4~~\text{with}~~b_2=\frac{19}{2},\label{betag2}\\
\frac{d\lambda_2}{dt'} &=& 30\lambda_2^2+\frac{45}{2}\lambda_4^2+51\lambda_2\lambda_4-144g_2^2\lambda_2,\label{betalam2}\\
\frac{d\lambda_4}{dt'} &=& 288g_2^4+\frac{255}{8}\lambda_4^2+24\lambda_2\lambda_4-144g_2^2\lambda_4.\label{betalam4}
\end{eqnarray}
We have neglected the $2\lambda_3^2$ term in Eq~\eqref{betalam2} but it would not affect the result a lot.

The things we would like to do is similar to the treatment in Ref.~\cite{Hamada:2015bra}, but we include the running effect of $g_2$ rather than treat $g_2$ as constant.
We assume that
\begin{eqnarray}
\lambda_2(t')=f_1(t')g_2^2(t'),~\lambda_4(t')=f_2(t')g_2^2(t'),
\end{eqnarray}
where $f_{1,2}(t')$ are some functions of $t'$. Substitute them into Eqs.~\eqref{betalam2}--\eqref{betalam4}, and we find
\begin{eqnarray}
&&\frac{df_1}{dG}=30f_1^2+\frac{45}{2}f_2^2+51f_1f_2-(144+2b_2)f_1,\label{betaf1}\\
&&\frac{df_2}{dG}=288+\frac{255}{8}f_2^2+24f_1f_2-(144+2b_2)f_2.\label{betaf2}
\end{eqnarray}
where $dG=g_2^2(t')dt'$. The function $G$ can be easily obtained by integration and the result is
\begin{eqnarray}
G(t')=\frac{1}{b_2}\ln\left(\frac{g_2(t')}{g_2(t'_0)}\right).
\end{eqnarray}
The right hand side of Eq.~\eqref{betaf1} does not contain a constant, but the right hand side of Eq.~\eqref{betaf2} does.

We treat these equations in a way similar to what was did in Ref.~\cite{Hamada:2015bra}. Firstly, redefine $f_2$ to remove the $24f_1f_2$ term. This can be done by linearly combining $f_1$ and $f_2$ as $F=f_1+\eta f_2$ with $\eta=(17+\sqrt{7968})/64\approx1.66046$. Then the equation of $F$ is
\begin{equation}
\frac{dF}{dG}=c_0-c_1F+c_2F^2+c'f_1^2,\label{betaF}
\end{equation}
where
\begin{eqnarray}
c_0&=&288\eta\approx478.212,~
c_1=144+2b_2=163,\\
c_2&=&\frac{12(153+\sqrt{7969})}{\sqrt{7969}+17}\approx27.3572,~
c'=30-c_2\approx2.64278.
\end{eqnarray}
The effect of $c'f_1^2$ in Eq.~\eqref{betaF} can be neglected because $c'$ is small, and we can analytically solve the equation. The solution is
\begin{equation}
F(t')=\frac{c_1}{2c_2}+\hat{d}\tan\left\{c_2\hat{d}G(t')+\tan^{-1}\left[\frac{1}{\hat{d}}\left(F(t'_0)-\frac{c_1}{2c_2}\right)\right]\right\},
\end{equation}
with
\begin{equation}
\hat{d}=\sqrt{\frac{c_0}{c_2}-\frac{c_1^2}{4c_2^2}}\approx2.93346.
\end{equation}
Here $t'_0$ is chosen to correspond to a scale $\Lambda_0$ where $y(t'_0)$ is small compared to $g_2(t'_0)$. We can see that $F(t')$ diverges when $$c_2\hat{d}G(t')+\tan^{-1}\left[\frac{1}{\hat{d}}\left(F(t'_0)-\frac{c_1}{2c_2}\right)\right]=\frac{\pi}{2}.$$
It means for any initial value of $F(t'_0)$, there always exists a Landau pole of $F(t')$ with a finite value of $G(t')$. This LP scale is
\begin{eqnarray}
\Lambda_\mathrm{LP}^{(F)}&=&\Lambda_0\exp\left\{\frac{8\pi^2}{b_2g_2^2(t_0)}\left[1-\exp\left(-\frac{b_2\pi}{c_2\hat{d}}\left[1-\frac{2}{\pi}\tan^{-1}\left(\frac{F(t'_0)}{\hat{d}}-\frac{c_1}{2c_2\hat{d}}\right)\right]\right)\right]\right\}\nonumber\\
&=&\Lambda_\mathrm{LP}^{(g_2)} \exp\left[-\frac{8\pi^2}{b_2g_2^2(t'_0)}\exp\left(-\frac{b_2\pi}{c_2\hat{d}}\left[1-\frac{2}{\pi}\tan^{-1}\left(\frac{F(t'_0)}{\hat{d}}-\frac{c_1}{2c_2\hat{d}}\right)\right]\right)\right].
\end{eqnarray}

\begin{figure}[!htbp]
\centering
\includegraphics[width=0.7\textwidth]{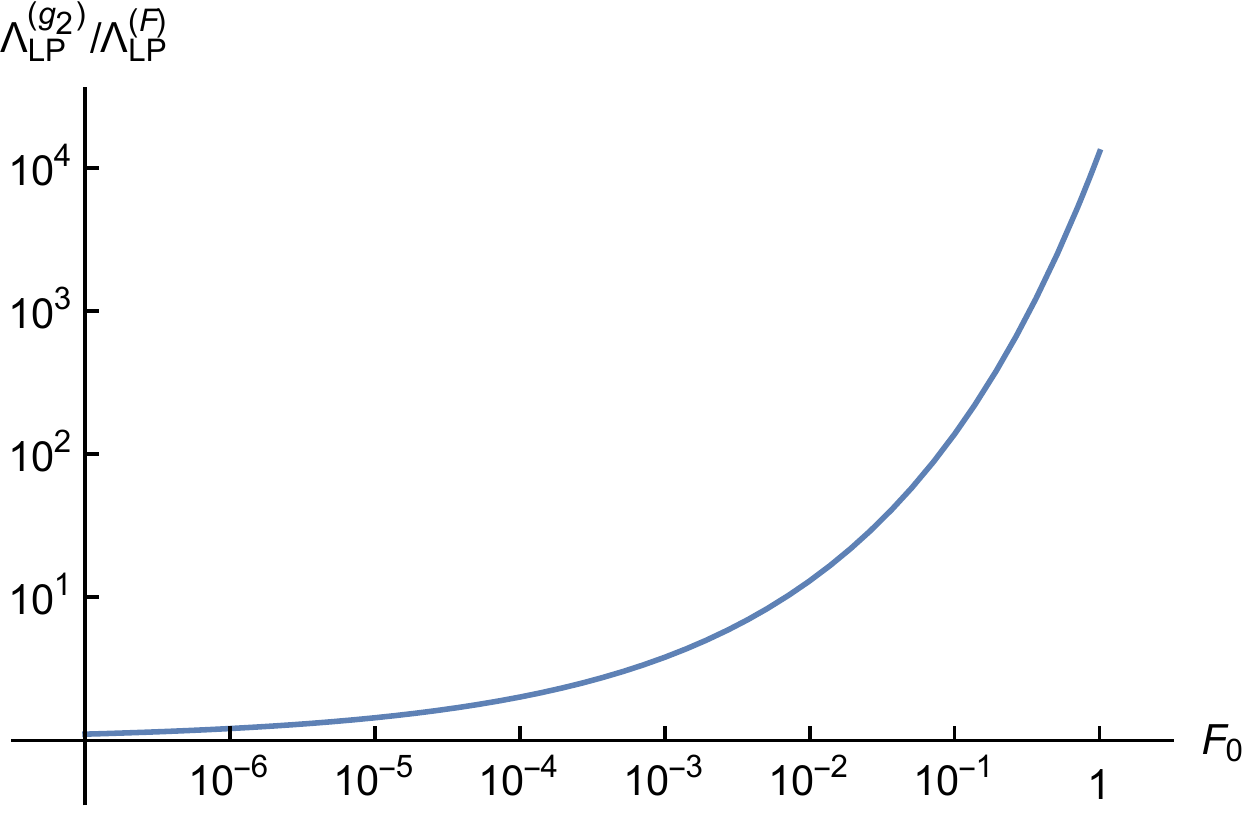}
\caption{The ratio $\Lambda_\mathrm{LP}^{(g_2)}/\Lambda_\mathrm{LP}^{(F)}$ as a function of ${F_0}$ in the range $0<{F_0}\leq1$.}\label{F0positive}
\end{figure}

If we choose the range of $y(0)$ satisfying $25y_c^2/26\leq y^2(0)<y_c^2$ to make $0<{F_0}\leq1$, we can assume that $y$ is small enough to be ignored in the running when the ratio
$$\xi\equiv\frac{y^2(t'_0)}{(24+b_2)g_2^2(t'_0)/25}=\left[1+\frac{{F_0}}{25}\left(\frac{g_2^2(t'_0)}{g_2^2(t'_f)}\right)^{(24+b_2)/b_2}\right]^{-1}$$
is small enough. This will fix the value of $g_2(t'_0)$ by
\begin{equation}
\frac{1}{g_2^2(t'_0)}=\frac{1}{g_2^2(0)}\left[\frac{{\xi F_0}}{25(1-\xi)}\right]^{b_2/(24+b_2)}.
\end{equation}
Notice that from the scale $M_{35}$ to the scale $\Lambda_0$, the function of $y(t)$ can be approximated by that in the ${F_0}=0$ limit. This means that in this scale range, the approximate solutions of $\lambda_i$ and $\lambda$ to Eqs.~\eqref{assumption1} work well if $\Lambda_0$ is high enough.  Then we can estimate $F(t'_0)$ by $a_1+\eta a_2\approx1.359$, and the LP scale is
\begin{eqnarray}
\Lambda_\mathrm{LP}^{(F)}&\approx& \Lambda_\mathrm{LP}^{(g_2)}\exp\Bigg[-\frac{8\pi^2}{b_2g_2^2(0)}\left(\frac{{\xi F_0}}{25(1-\xi)}\right)^{b_2/(24+b_2)}\nonumber\\*
&&\qquad\qquad\qquad\times\exp\left(-\frac{b_2\pi}{c_2\hat{d}}\left[1-\frac{2}{\pi}\tan^{-1}\left(\frac{a_1+\eta a_2-c_1/(2c_2)}{\hat{d}}\right)\right]\right)\Bigg].\qquad
\end{eqnarray}
For $F_0\leq1$, $\xi=0.895$ leads to an agreement with the full numerical calculation. We show the ratio $\Lambda_\mathrm{LP}^{(g_2)}/\Lambda_\mathrm{LP}^{(F)}$ as a function of $F_0$ in Fig.~\ref{F0positive}. We can see that for $F_0=1$, the LP scale of $F$ is smaller than $\Lambda_\mathrm{LP}^{(g_2)}$ by 4 orders of magnitude. If one wants $\Lambda_\mathrm{LP}^{(F)}$ to be $\sim 10^{-1}\Lambda_\mathrm{LP}^{(g_2)}$, $F_0$ should be tuned to $\sim 0.002$, which is unnatural. For a larger ${F_0}$, $\Lambda_\mathrm{LP}^{(F)}$ is lower, because the 7-3-5 Yukawa coupling has smaller effect on the running of quartic couplings.

\section{Conclusions and discussions}\label{sect4}

Perturbativity yields strong bounds on the MDM model, but most of the previous studies only consider the gauge couplings and conclude that the real scalar septuplet MDM model can keep perturbative up to the Planck scale. In this article we take into account the quartic self-interactions of the real scalar septuplet in the septuplet MDM model, and find that for an MDM particle mass of $\sim10$~TeV the Landau poles of the quartic couplings appear around $10^8$~GeV, which is consistent with the approximate analytical treatment in Ref.~\cite{Hamada:2015bra}.

As an attempt to push up the LP scale, we propose an extension to the model with Yukawa interactions among the septuplet, an extra fermionic quintuplet, and one or two extra fermionic triplets. In principle, the Landau pole can be deferred to appear at $10^{14}$~GeV, but it is at the price of a serious fine-tuning of the initial condition of the new Yukawa coupling. We investigate the evolution of couplings up to a scale just a little bit below the highest possible LP scale, and use the VS and perturbativity conditions to constrain the quartic couplings. It is found that the Higgs portal coupling has an acceptable range of $-0.25<\lambda_3<1$. Moreover, if the MDM particle mass could be relaxed to $\sim10^8$~GeV, which demands some nonthermal production mechanisms, we could push up the LP scale even beyond the Planck scale.

In a scalar MDM model with doublet, triplet, or quadruplet, the couplings can remain perturbative up to the Planck scale.
On the other hand, when the MDM scalar multiplet lives in an $SU(2)_L$ representation with a dimension higher than 5, the scalar MDM model would have Landau poles appearing before the Planck scale, due to the MDM framework, i.e., the MDM particle mass fixed by thermal freeze-out dynamics. As what we propose in this article, these models may be cured as well by introducing extra Yukawa couplings. We leave this possibility to a specific study~\cite{five}.

\section*{Acknowledgments}

This work is supported by the National Natural Science Foundation of China (NSFC) under Grant Nos.~11375277, 11410301005, and 11005163,
the Fundamental Research Funds for the Central Universities, and the Sun Yat-Sen University Science Foundation.

\end{document}